\documentclass[%
 reprint,
 superscriptaddress,
 amsmath,amssymb,
 aps,
floatfix,
longbibliography]{revtex4-1}

\usepackage{graphicx,subcaption}
\usepackage{bm}

\usepackage{xcolor}
\usepackage[colorlinks=true,citecolor=blue,linkcolor=magenta]{hyperref}

\usepackage{verbatim}

\usepackage{multirow}
\usepackage{array}

\newcommand{\nn}{\nonumber}

\newcommand{\rar}{\rightarrow}

\newcommand{\ben}{\begin{eqnarray}}
\newcommand{\een}{\end{eqnarray}}

\newcommand{\be}{\begin{equation}}
\newcommand{\ee}{\end{equation}}

\begin{document}
\title{Graph-theoretical search for integrable 
multistate Landau-Zener models}

\author{Zixuan Li}
\affiliation{School of Physics and Electronics, Hunan University, Changsha 410082, China}
\author{Chen Sun}
\email{chensun@hnu.edu.cn}
\affiliation{School of Physics and Electronics, Hunan University, Changsha 410082, China}

\begin{abstract}
The search for exactly solvable models is an evergreen topic in theoretical physics. 
In the context of 
multistate Landau-Zener models---$N$-state quantum systems with linearly time-dependent Hamiltonians---the theory of integrability provides a framework for 
identifying new solvable cases. In particular, it was proved that the integrability of a specific class known as the multitime Landau-Zener (MTLZ) models guarantees their exact solvability. 
A key finding was that an $N$-state MTLZ model can be 
represented by data defined on an $N$-vertex graph. While known host graphs for MTLZ models include hypercubes, fans, and their Cartesian products, no other families have been discovered, leading to the conjecture that these are the only possibilities. In this work, we conduct a systematic graph-theoretical search for integrable models within the MTLZ class. By first identifying minimal structures that a graph must contain to host an MTLZ model, we formulate an efficient algorithm to systematically search for candidate graphs for MTLZ models. Implementing this algorithm using computational software, we enumerate all candidate graphs with up to $N = 13$ vertices and perform an in-depth analysis of those with $N \le 11$. Our results corroborate the aforementioned conjecture for graphs up to $11$ vertices. For even larger graphs, we propose a specific family, termed descendants of ``$(0,2)$-graphs'', as promising candidates that may violate the conjecture above. 
Our work can serve as a guideline to identify new exactly solvable multistate Landau-Zener models in the future.

\end{abstract}

\maketitle

\section{Introduction}

The pursuit of exactly solvable quantum models has been a central and enduring theme in quantum physics. The value of these ``beautiful models'' (as eulogized in the title of Sutherland's book \cite{Sutherland-2004}) is manifold: they enable accurate descriptions of physical systems through their exact solutions, serve as benchmarks for other unsolvable models, and provide platforms for exploring the symmetries and mathematical structures underlying exact solvability. In the domain of time-dependent quantum problems, exactly solvable models have been found since the birth of quantum mechanics---in four pioneering works in 1932, Landau \cite{landau}, Zener \cite{zener}, St\"{u}ckelberg \cite{stuckelberg}, and Majorana \cite{majorana} independently solved a fundamental two-state model with a Hamiltonian linear in time, now known as the Landau-Zener (LZ) model \cite{Shevchenko-2010,Shevchenko-2023}. Since then, a variety of exactly solvable \cite{note-solvable} multistate generalizations of the LZ model---models with Hamiltonians of the form $H=Bt+A$, where $A$ and $B$ are $N\times N$ constant matrices and $N\ge 3$)---have been identified \cite{Wei-1963,DO,Hioe-1987,bow-tie,Rau-1998,GBT-Demkov-2000,GBT-Demkov-2001,chain-2002,4-state-2002,Rau-2003,Rau-2005,Vasilev-2007,chain-2013,Patra-2015,DTCM-2016,DTCM-2016-2,HC-2017,cross-2017}. Methods for exactly solving the original LZ and these multistate LZ (MLZ) models typically involve sophisticated mathematical techniques, such as special functions \cite{zener}, Laplace transformations \cite{majorana,DO,bow-tie,GBT-Demkov-2001}, Lie-algebraic methods \cite{Wei-1963,Hioe-1987,Rau-1998,Rau-2003,Rau-2005,Patra-2015
}, and analytical constraints \cite{HC-2017,cross-2017}.

The recently developed theory of integrability for time-dependent quantum systems \cite{Patra-2015,6-state-2015,4-state-2015,quest-2017,commute} sheds new light on the search for solvable MLZ models. According to \cite{commute}, a Hermitian time-dependent quantum Hamiltonian $H$ (an $N\times N$ matrix) is 
{\it integrable} if there exists another Hermitian operator $H'$ (also an $N\times N$ matrix) such that the following ``zero curvature conditions'' are satisfied:
\begin{align}\label{int-cond}
[H,H']=0,\quad \frac{\partial H}{\partial \tau }=\frac{\partial H'}{\partial t },
\end{align}
where $\tau$ is another parameter (besides the time $t$) that $H$ and $H'$ depend on. We refer to $H'$ as a ``commuting partner'' of $H$. Note that $H'$ needs to be nontrivial---it cannot be merely a linear combination of $H$ and the identity matrix. The conditions \eqref{int-cond} guarantee that the nonabelian gauge field defined as ${\mathcal A}(t, \tau) = -i(H,H')$ has zero curvature. As a consequence, the original evolution path along time $t$ (namely, from $t=-\infty$ to $t=\infty$ at a fixed $\tau$) can be deformed in the $(t,\tau)$ plane without changing the evolution operator (provided that the deformation does not cross any singularities of $H$ or $H'$):

\begin{align}\label{U-LZ-deform}
\mathcal{T} e^{-i\int_{-\infty}^\infty H dt}=\mathcal{T_\mathcal{P}} e^{-i\int_{\mathcal{P} } (H dt+ H'd\tau)},
\end{align}
where $\mathcal{P}$ is a path in the $(t,\tau)$ plane with the same starting and ending points as the original evolution, and $\mathcal{T_\mathcal{P}}$ is the ordering operator defined along the path $\mathcal{P}$ \cite{footnote-complex-analogy}. For a specific Hamiltonian that satisfies the integrability conditions, this freedom of path deformation serves as a powerful tool to simplify the evolution problem. 
Indeed, integrability has been utilized to obtain exact solutions or approximate semiclassical solutions of many different types of MLZ models \cite{Yuzbashyan-2018,large-2018,gamma-2019,MTLZ-2020,parallel-2020,quadratic-2021,Malla-2021} and of their generalizations to non-Hermitian systems \cite{Malla-2023}, to systems with time-dependencies beyond linear ones \cite{Li-2018,Barik-2025}, and to classical systems where the commutation relation in \eqref{int-cond} is replaced by a Poisson bracket \cite{Tyagi-2025}.

\begin{figure}[!htb]
\scalebox{0.65}[0.65]{\includegraphics{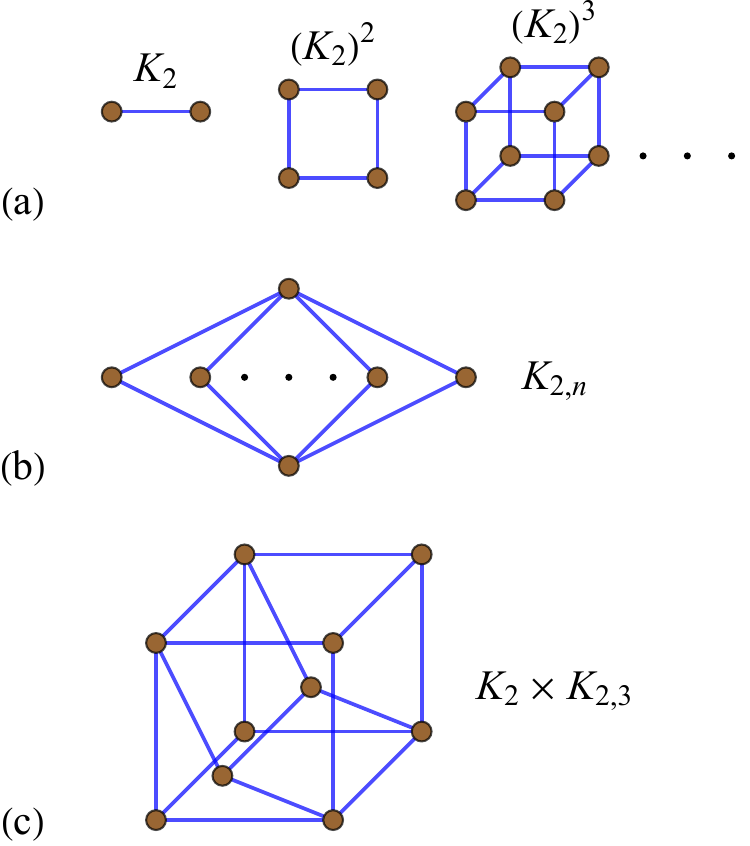}}
\caption{Examples of known of graphs that host MTLZ models: (a) the hypercube graphs $Q_D\equiv(K_2)^D$ with $D\ge 1 $; (b) the fan graphs $K_{2,n}$ with $n\ge 3$; (c) a Cartesian product of a $K_{2}$ and $K_{2,3}$. }
	\label{fig:solvable-graphs}
\end{figure}

It is worth emphasizing that integrability (as defined in \cite{commute}) of an MLZ model does not always lead to its exact solvability, as illustrated by models considered in \cite{quadratic-2021,Malla-2021,Hu-5-state}. However, for one large class of MLZ models, integrability does guarantee exact solvability. This class is characterized by the existence of a commuting partner $H'$ such that both $H$ and $H'$ depend linearly on $t$ and $\tau$, namely, $H=B_{11} t+B_{21} \tau+A_1$ and $H'=B_{12} t+B_{22} \tau+A_2$ \cite{MTLZ-2020,large-2018}. 
Therefore, this class serves as an ideal playground for systematic search for solvable MLZ models. Models within this class are named {\it multitime} Landau-Zener (MTLZ) models, because $\tau$ plays similar roles as the original time $t$. In \cite{MTLZ-2020}, it was demonstrated that for an $N$-state MTLZ model, the integrability conditions \eqref{int-cond} reduce to a set of constraints which can be conveniently presented as data on an $N$-vertex {\it graph}, thus allowing systematic classification of MTLZ models by graphs. In particular, the  ``hypercube'' family of graphs (Fig.~\ref{fig:solvable-graphs}(a)) and the ``fan'' graphs (Fig.~\ref{fig:solvable-graphs}(b)) are shown to host MTLZ models \cite{MTLZ-2020}. Besides, models made of direct products of MTLZ models are automatically MTLZ models as well, which means that Cartesian products (defined later) of hypercubes and fans are also graphs hosting MTLZ models. In \cite{nogo-2022}, additional necessary conditions for graphs that may host MTLZ models are proved in the forms of two ``no-go rules'', which further facilitates the search for solvable models in the MTLZ class; all possible graphs with no more than $9$ vertices are analyzed, and it was shown that there are no more graphs besides the hypercubes and the fans that host MTLZ models. By induction, a natural conjecture is to extend this to graphs of any size:

\smallskip
{\it \noindent {\bf Conjecture:}  
The hypercubes, the fans and their Cartesian products are the only graphs to host MTLZ models}.
\smallskip

\noindent To prove this conjecture one should make some general argument to exclude all other graphs, which has not been achieved so far; whereas to disprove it one simply needs to find a counterexample in a larger graph beyond $9$ vertex ones. 
However, the analysis of larger graphs is also hampered by a drastic growth in complexity, stemming from both the increasing number of possible graphs and the increasing complexity of each individual graph.

In this work, motivated by both the search for new exactly solvable models and the desire to address the aforementioned conjecture, we perform a systematic graph-theoretical search for MTLZ models
. Starting from identifying minimal structures that a graph must contain to host MTLZ models, we develop an algorithm to search for candidate graphs for MTLZ models that can be conveniently implemented on computational software. We further perform analysis on each candidate graph with $10$ and $11$ vertices. Our results indicate that the conjecture above, although still not completely resolved, is valid up to graphs with $11$ vertices. We also propose to investigate certain classes of even larger graphs which may violate the conjecture above. 

This work is organized as follows. In Sec. II, we review known results for MTLZ models and their representations on graphs
. In Sec. III, general graph-theoretical analyses are performed to construct minimal structures for candidate graphs for MTLZ models. Based on these analyses, in Sec. IV an algorithm to search for candidate graphs is developed and implemented on computational software to identify all candidate graphs with no more than $13$ vertices, and detailed analysis are performed on each candidate graph with $10$ and $11$ vertices. In Sec. V, several graphs with more than $13$ vertices, which are descendants of the so-called $(0,2)$-graphs, are analyzed; we expect that they are promising candidates to break the conjecture above. Sec. VI contains a brief discussion on non-bipartite graphs. Finally, Sec. VII presents conclusions and discussions.

\section{Basics of multitime Landau-Zener models}

In this section, we review known results for MTLZ models with a focus of their representation on graphs. For readers interested in knowing more details about MTLZ models, we refer to \cite{MTLZ-2020}. 


\subsection{Notations}

We first establish the notation used throughout this paper.

We are going to follow standard notations in graph theory, for example those in Diestel's book \cite{Diestel}; here we review the most basic definitions, and put some other definitions as footnotes at their first appearances. A {\it graph} $G$ is defined as a pair of sets $(V,E)$, where $V$ consists of {\it vertices} $a$ and $E$ consists of {\it edges} $ab$ connecting vertices $a$ and $b$. All graphs are assumed to be simple, namely without parallel edges or loops (edge with both ends at the same vertex). Two vertices are {\it adjacent} if there is an edge between them. A graph is {\it complete} if all its vertices are pairwise adjacent. A complete graph with $m$ vertices is denoted by $K_{m}$. A graph is {\it bipartite} if its vertices can be separated into two groups such that any two vertices in the same group are not adjacent. A bipartite graph is {\it complete bipartite} if any two vertices in the two different groups are adjacent. A complete bipartite graph is denoted by $K_{m,n}$, where $m$ and $n$ are numbers of vertices in the two groups.

We will also employ some shorthand notations for simplicity. In particular, we will call a path or a cycle of length $l$ simply as an ``{\it$l$-path/cycle}'', for example, a $2$-path, a $4$-cycle, etc. For two vertices $a$ and $b$, we will write $a\sim b$ if they are adjacent, and $a\nsim b$ if they are not adjacent. Sequences of vertices connected by $\sim$ will be used to denote paths or cycles, e.g. $a\sim b\sim c$. 
A graph $G$ is said to be ``{\it$H$-free}'' (with $H$ being another graph) if $G$ does not contain $H$ as a subgraph. We will call the graph shown in Fig.~\ref{fig:K33and1221}(b) an ``{\it$1221$ graph}''. It can be obtained by deleting an edge from the graph $K_{3,3}$ shown in Fig.~\ref{fig:K33and1221}(a). $G_1 \times G_2$ means the {\it Cartesian product} of two graphs $G_1$ and $G_2$ (see footnote \cite{footnote-Cartesian} for its definition); for example, Fig.~\ref{fig:solvable-graphs}(c) shows the Cartesian product of $K_2$ and $K_{2,3}$. A power of a graph [e.g. $(K_2)^3$ and $(K_2)^3$ in Fig.~\ref{fig:solvable-graphs}(a)] is understood in the Cartesian product sense.

\subsection{Multitime Landau-Zener models on graphs}

In this subsection, we review how MTLZ models are represented as data defined on graphs.

To a given $N$-state MLZ model, we associate a 
graph $G = (V, E)$ with $N$ vertices. Each vertex $a\in V$ corresponds to a diabatic state labeled by $a$
, and each edge $ab \in  E$ corresponds to a {\it nonzero coupling} between the diabatic states $a$ and $b$. On each edge $ab$ one defines three types of data: a sign $s^{ab}=-s^{ba}$, a linear form $\bar A^{ab}=\bar A_{j}^{ab} dx^{j}$, and an antisymmetric parameter $\gamma^{ab}=-\gamma^{ba}$. In \cite{MTLZ-2020}, it was proved from the integrability conditions \eqref{int-cond} that the necessary and sufficient condition for such a graph to host an MTLZ model is that there exist nontrivial solutions for $s^{ab}$, $\bar A^{ab}$ and $\gamma^{ab}$ to a set of equations obtained from the following two properties:

\smallskip
{\bf  \noindent  1.~Cycle property:}
For any cycle $a_1\sim a_2 \sim \ldots\sim  a_k \sim a_1$ in the graph
\begin{eqnarray}
\label{cycle-bar-A} \sum_{l=1}^k s_{a_l,a_{l+1}} \bar{A}^{a_l,a_{l+1}} \otimes \bar{A}^{a_l,a_{l+1}} = 0,
\end{eqnarray}
where ``$\otimes$'' denotes the tensor direct product, and the index $a_{k+1}$ appeared in the summation should be understood as $a_1$.

{\bf  \noindent  2.~Multipath property:}
For any pair of vertices $a$ and $b$ with a {\it distance} $2$ in the graph \cite{note-distance}
\begin{eqnarray}
\label{multipath-bar-A} \sum_{c} \sqrt{|\gamma^{ac}\gamma^{bc}|} \bar{A}^{ac} \wedge \bar{A}^{bc} = 0,
\end{eqnarray}
where ``$\wedge$'' denotes the skew symmetric tensor product (the wedge product), and the summation goes over all $2$-paths $a\sim c \sim b$ in the graph that connect $a$ and $b$. Nontrivial solutions mean that the $\bar{A}^{ab}$ on any two adjacent edges needs to be linearly independent, namely, $\bar{A}^{ac} \wedge \bar{A}^{bc} \ne 0$, and all $\gamma^{ab}$'s needs to be non-zero. Note that the number of independent equations from the cycle property is equal the number of independent cycles in a graph. 

\begin{figure}[!htb]
(a)~ \scalebox{0.25}[0.25]{\includegraphics{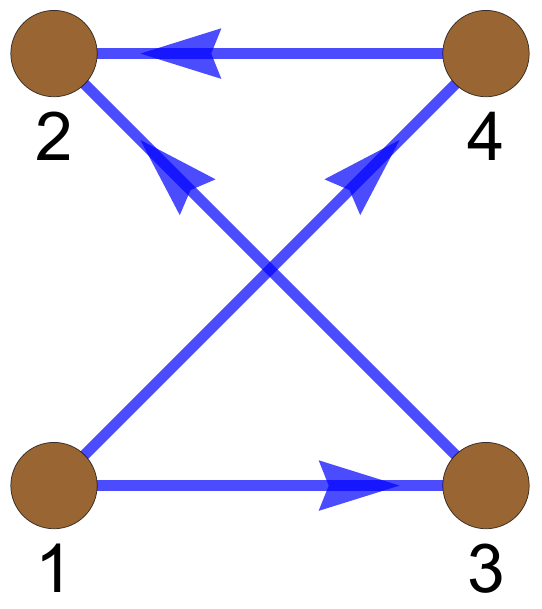}} ~ ~ ~ ~
(b)~ \scalebox{0.25}[0.25]{\includegraphics{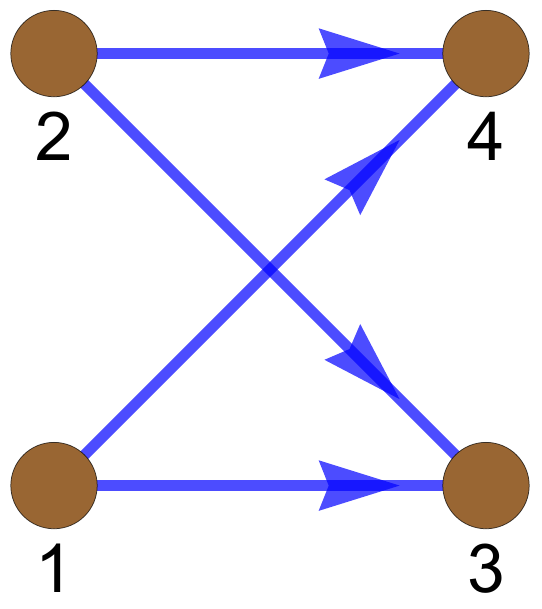}}
\caption{
Graphs of a $4$-cycle with two types of orientations: (a) the non-bipartite orientation; (b) the bipartite orientation.}
\label{fig:4-loop-graph}
\end{figure}

The two properties further determine relations on all $4$-cycles in a graph. The signs $s^{ab}$ can be represented by adding arrows on edges in the original graph, so it becomes a {\it directed} graph: namely, one draws $a\rar b$ if $s^{ab}=-1$, and $b \rar a$ if $s^{ab}=1$. Then the rules for the signs $s^{ab}$ in any $4$-cycles $1\sim3\sim2\sim4\sim1$ are such that it must be either of the ``non-bipartite orientation'' shown in Fig.~\ref{fig:4-loop-graph}(a), or of the ``bipartite orientation'' shown in Fig.~\ref{fig:4-loop-graph}(b).

For the non-bipartite orientation in Fig.~\ref{fig:4-loop-graph}(a), the $\bar A^{ab}$ forms are related by a pseudo-orthogonal transformation:
\begin{eqnarray}
\label{A-LO-transf-correspond-2} \bar{A}^{24} &=& p \left(\cosh\vartheta \bar{A}^{13} - r \sinh\vartheta \bar{A}^{14}\right), \nonumber \\ \bar{A}^{23} &=&
p\left(\sinh\vartheta \bar{A}^{13} - r \cosh\vartheta \bar{A}^{14}\right),
\end{eqnarray}
where $p=\pm 1$ and $r=\pm 1$ are sign factors, and $\vartheta$ is a real parameter called a ``rapidity''. From this follow two relations of wedge products:
\begin{eqnarray}
&\label{A-LO-determinants-tilde} \bar{A}^{23} \wedge \bar{A}^{24} = r \bar{A}^{13} \wedge \bar{A}^{14},\nn\\
&\bar{A}^{14} \wedge \bar{A}^{24} = r \bar{A}^{13} \wedge \bar{A}^{23}.
\end{eqnarray}
For the bipartite orientation in Fig.~\ref{fig:4-loop-graph}(b), the $\bar A^{ab}$ forms are related by a different pseudo-orthogonal transformation:
\begin{eqnarray}
\label{A-LO-transf} \bar{A}^{24} &=& p \left( r \sinh\vartheta \bar{A}^{13}+\cosh\vartheta \bar{A}^{14}\right), \nonumber \\ \bar{A}^{23} &=& p\left( r \cosh\vartheta
\bar{A}^{13}+\sinh\vartheta \bar{A}^{14} \right).
\end{eqnarray}
From this follows two relations of wedge products:
\begin{eqnarray}
\label{connect-bipart} &\bar{A}^{23} \wedge \bar{A}^{24} = r \bar{A}^{13} \wedge \bar{A}^{14},\nn\\
&\bar{A}^{14} \wedge \bar{A}^{24} =- r \bar{A}^{13} \wedge \bar{A}^{23}.
\end{eqnarray}
Note that the signs for these two wedge product relations are different for the bipartite orientation, whereas those signs are the same for the non-bipartite orientation. With the wedge product relations, the multipath property \eqref{multipath-bar-A} becomes a condition on $\gamma^{ab}$:
\begin{eqnarray}
\label{multipath-gamma} \sqrt{|\gamma^{ad}\gamma^{bd}|}+ \sum_c^{c\ne d} r_{acbd}  \sqrt{|\gamma^{ac}\gamma^{bc}|} = 0,
\end{eqnarray}
where $r_{acbd}$ is the $r$ factor appeared in the wedge product relation $\bar{A}^{ac} \wedge \bar{A}^{bc} = r_{acbd} \bar{A}^{ad} \wedge \bar{A}^{bd}$.

To identify an MTLZ solution, one first specifies the sign $s^{ab}$ by drawing the graphs as a directed graph. Then one should find all possible sets of the $r$ factors on all $4$-cycles. Then one should determine if there is solution to the system of equations of $\bar A^{ab}$ for all $4$-cycles, and if there is a solution to the system of equations of $\sqrt{|\gamma^{ab}|}$ for all pairs of vertices at distance $2$. If there are no solutions at any specific step, one can conclude that the directed graph admits no solutions and terminate the analysis. For example, there may be simply no solution of $r$ factors in a directed graph, in which case this directed graph does not host MTLZ models.

For simplicity, we will call a graph that hosts MTLZ models an ``{\it integrable graph}''.

\subsection{Rules for integrable graphs}

The graph presentation of MTLZ models in principle enables a systematic search for MTLZ models by considering different graphs. Since the number of graphs increases exponentially at increasing number of vertices, one may expect that this searching task quickly becomes formidable at increasing number of vertices. Fortunately, the two properties \eqref{cycle-bar-A} and \eqref{multipath-bar-A} dictates that integrable graphs must obey strict rules that considerably reduce the number of graphs that truly need to be considered. Specifically, let $G$ be an integrable graph, then it must satisfy the following four rules proved in \cite{MTLZ-2020,nogo-2022}:

\smallskip
{\bf   \noindent 1.~No $K_3$ rule:} $G$ is $K_3$-
free.

{\bf  \noindent  2.~$2$-path rule:} For any pair of vertices in $G$ that has distance $2$, there must be at least two $2$-paths between these two vertices.

\begin{figure}[!htb]
(a)~ \scalebox{0.25}[0.25]{\includegraphics{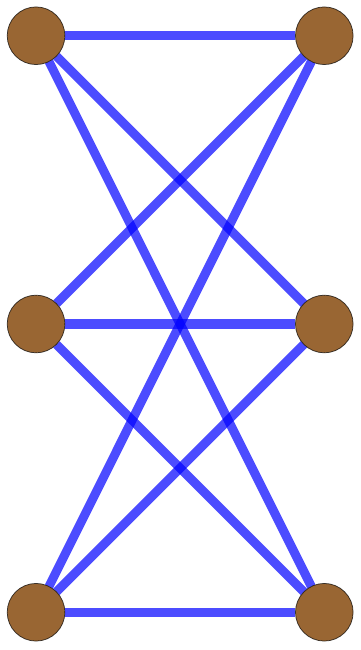}} ~ ~ ~
(b)~ \scalebox{0.42}[0.42]{\includegraphics{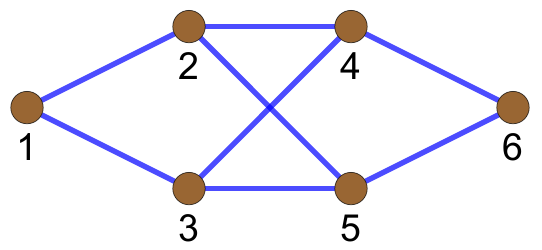}}
\caption{
Graphs appeared in the description of the rules for candidate graphs: (a) the complete bipartite graph $K_{3,3}$; (b) the $1221$ graph.}
\label{fig:K33and1221}
\end{figure}

{\bf  \noindent  3.~No $K_{3,3}$ rule:} $G$ is $K_{3,3}$-free ($K_{3,3}$ is shown in Fig.~\ref{fig:K33and1221}(a)).

{\bf  \noindent  4.~No $1221$ rule:} If $G$ contains a $1221$ graph as a subgraph with vertices labelled as in Fig.~\ref{fig:K33and1221}(b), there must exist another vertex $a$ in $G$ such that at least one of the four $2$-paths $1\sim a\sim 4$, $1\sim a\sim 5$, $2\sim a\sim 6$ or $3\sim a\sim 6$ exists. 
\smallskip

\noindent 
These rules are necessary conditions for integrable graphs---if any condition is violated in a graph, it is guaranteed to host no MTLZ models; but they are not sufficient conditions. Hence, we will call a graph that satisfies all the four rules a ``{\it candidate graph}''; further analysis is required to determine if it is an integrable graph.

Rules 1 and 3 are relatively easy to understand---they simply mean that $K_3$ (triangle) and $K_{3,3}$ are forbidden as subgraphs. Rules 2 and 4 need more explanation. According to Rule 2, if there is a $2$-path $a\sim c \sim b$ in $G$, then there must exist another vertex $d$ such that $a\sim d\sim b$ in $G$. The vertices $a$, $b$, $c$ and $d$ thus form a $4$-cycle. Rule 4 does not mean that the $1221$ graph is forbidden as a subgraph; it instead means that if a $1221$ subgraph exists, there must also exist another $2$-path (involving another vertex) as described in the statement of the rule. Also note that the first two rules together indicate that $4$-cycles are elementary bricks for a candidate graph, whose properties were reviewed in the previous subsection.

The no $K_3$ rule and the $2$-path were proved in \cite{MTLZ-2020}. The $2$-path rule can be understood intuitively as the following. The multipath property \eqref{multipath-bar-A} is equivalent to the condition that the commutators $[A_j,A_k]$ vanish, where $A_{j(k)}$ are matrices with elements $\sqrt{|\gamma^{ab}|}A^{ab}_{j(k)}$. In those commutators, each pair of states $a$ and $b$ corresponds to an off-diagonal element, and in the associated graph each $2$-path between the vertices $a$ and $b$ contributes a non-zero term to that element. If there is just a single term, that element cannot be zero, hence the commutator cannot vanish. At least two terms (and thus two $2$-paths) are required to make that element zero, and their contributions should differ by a sign---namely, destructive interference must occur. Applying this argument to every pair of vertices of distance $2$ gives the $2$-path rule. The no $K_3$ rule can be explained by the cycle property \eqref{cycle-bar-A} and the requirement of linear-independence of adjacent $\bar A$ forms \cite{MTLZ-2020}. If there exists a $K_3$ subgraph with vertices $1$, $2$, and $3$, then the cycle property for the $3$-cycle $1 \sim 2 \sim 3 \sim 1$ reads $s_{12} \bar A^{12} \otimes \bar A^{12}+ s_{23} \bar A^{23} \otimes \bar A^{23} + s_{31} \bar A^{13} \otimes \bar A^{13} = 0$. Assuming that $\bar A^{13}$ is a linear superposition of the other two: $\bar A^{13} = \alpha \bar A^{12} +\beta \bar A^{23}$, one can check that this equation cannot be satisfied by any choices of coefficients $\alpha$ and $\beta$ and signs $s_{12}, s_{23}, s_{31}$, so this $K_3$ subgraph violates the cycle property. The no $K_{3,3}$ and no $1221$ rules seem to have no intuitive explanation, as they follow from detailed considerations on the solutions on a $4$-cycle \cite{nogo-2022}.

\section{Graph-theoretical analysis}

In this section, we perform general graph-theoretical analysis based on the four rules stated near the end of the previous section. 

We will assume any considered graph to be connected, finite, and bipartite. The assumption of connectedness is because for a disconnected graph, its corresponding MLZ model can always be decomposed into smaller MLZ models without interactions between them. The assumption of finiteness is because we focus on MLZ models with finite number of levels. Non-bipartite graphs are excluded for the convenience of theoretical analysis; they will be discussed in Sec. VI.


We will adopt the {\it layer graph} notation and the scheme for graph classification proposed in \cite{nogo-2022}, which we describe below. We denote by $N$ the number of vertices of a graph
. For a fixed $N$, we first choose the {\it diameter} $d$ of a graph \cite{note-diameter}. As argued in \cite{nogo-2022}, any bipartite graph with diameter $d$ can be drawn as a layer graph with $d+1$ layers---we call them the $0$th layer, the $1$st layer to the $d$th layer (the $0$th and the $d$th layers are ``outer'' layers, whereas all other layers are ``inner'' layers); 
vertices are distributed among these layers, and edges are allowed to exist only between vertices in adjacent layers (see Fig.~\ref{fig:innerlayer3} for examples of layer graphs). Enumeration of possible layer graphs at fixed $N$ and $d$ takes two steps. First, one fixes the numbers of vertices in each layer, which we denote as $N_0, N_1,\ldots, N_d$ for the $0$th, the $1$st up to the $d$th layer, with $N_0+N_1+ \ldots+ N_d=N$. Second, one considers all possible ways that edges may be drawn between adjacent layers. 

It has been shown in \cite{nogo-2022} that a layer graph constructed in this way is bipartite, and that a bipartite graph can always be drawn as a layer graph. Also note that one layer graph may be isomorphic 
to another layer graph with a different arrangement of vertices in the layers, and one is free to choose any graph in this set of equivalent graphs for further analysis. 


The no $K_3$ rule is automatically satisfied for a layer graph (bipartite graphs cannot have odd cycles \cite{Diestel}), 
and other rules are also relatively easy to be implemented on a layer graph; that is the reason we use this layer graph notation.

\subsection{Applications of the rules on graphs of different diameters}

We now apply the four rules to derive general properties of candidate graphs of different diameters $d$, starting from the smallest one.

1.~$d=1$:~Only the graph $K_2$ is possible, which corresponds to the $2$-state LZ model.

This follows immediately from the fact that a $d=1$ graph must be complete, and the only $K_3$-free complete graph is $K_2$. Note that $K_2$ has no $2$-path, so the $2$-path rule is satisfied.

2.~$d=2$:~Only complete bipartite graphs $K_{2,N-2}$ with $N\ge 4$ are possible, which correspond to the square (a hypercube at dimension $2$) model for $N=4$ and the fan models for $N\ge 5$.

The argument proceeds as follows. First, any $d=2$ candidate graph must be a complete bipartite graph. Assume that it is not a complete bipartite graph, we can show that its diameter $d$ must be at least $3$: take a pair of vertices in the two partitions that are not connected by an edge, then a shortest path between them is at least of length $2$, but its length cannot be $2$ since the ends of a $2$-path in a bipartite graph must be in the same partition; so such a shortest path must be at least of length $3$, and $d\ge 3$. Second, any complete bipartite graph $K_{m,n}$ with $m\ge 3$ and $n\ge3$ will have $K_{3,3}$ as a subgraph, which breaks the no $K_{3,3}$ rule. So only graphs $K_{2,n}$ are possible.

3.~$d\ge 3$:~Any inner layer must contain at least $3$ vertices.

\begin{figure}[!htb]
(a)~ \scalebox{0.25}[0.25]{\includegraphics{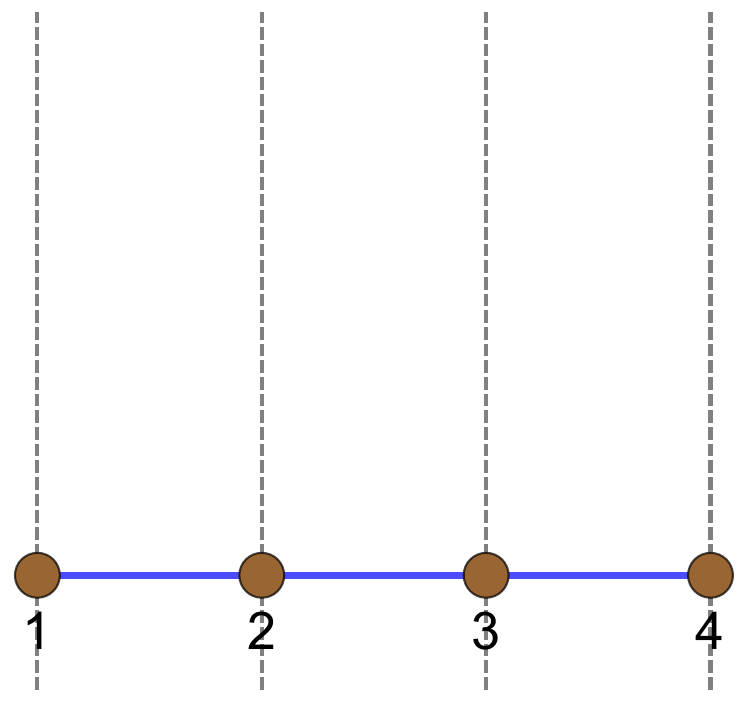}}
(b)~   \scalebox{0.25}[0.25]{\includegraphics{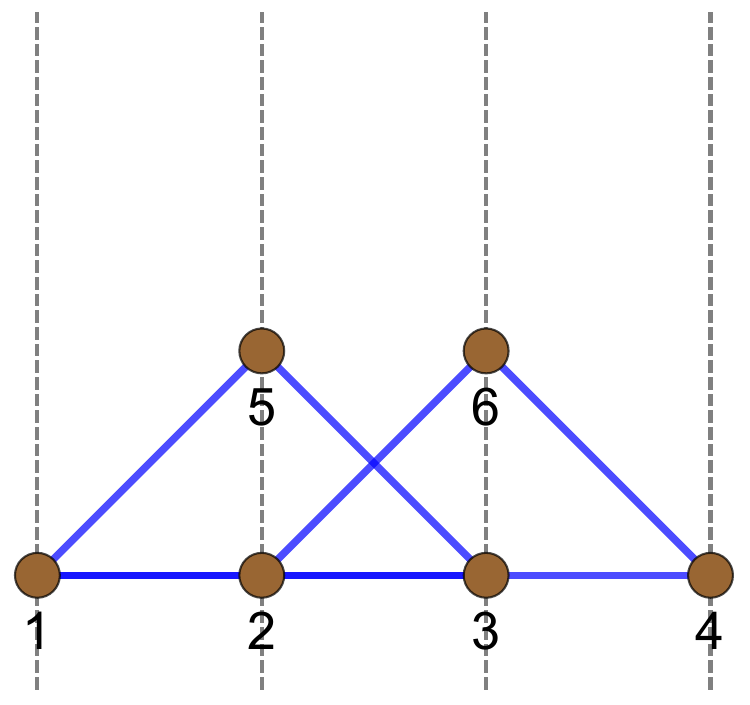}}
(c)~   \scalebox{0.25}[0.25]{\includegraphics{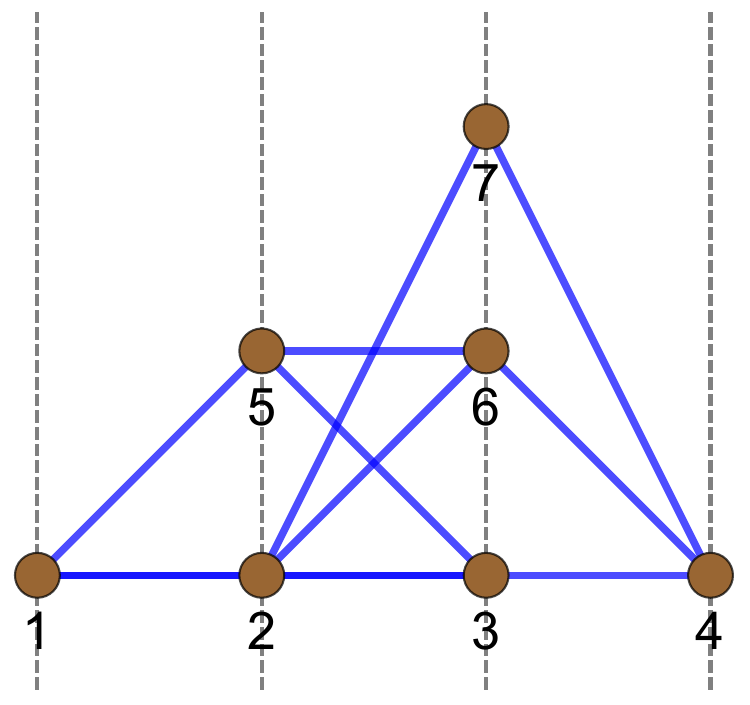}}
(d)~   \scalebox{0.25}[0.25]{\includegraphics{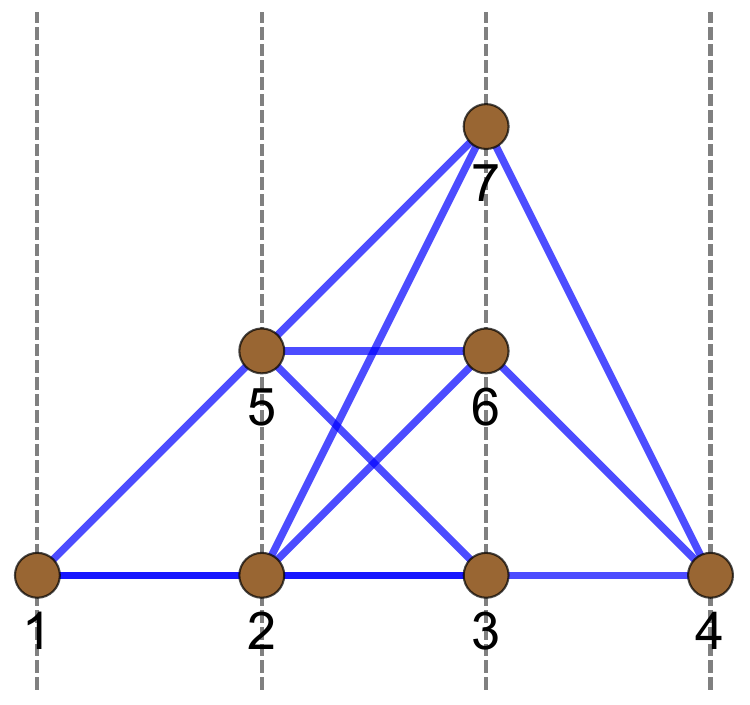}}
\caption{Arguments showing that for a graph with $d\ge 3$, any inner layer  must contain at least $3$ vertices. The dashed lines present different layers.}
\label{fig:innerlayer3}
\end{figure}

The situation for $d\ge 3$ is more complicated than the $d=1$ and $d=2$ cases. The argument for the above assertion proceeds as follows. Since the graph has diameter $d$, it must contain at least one $d$-path. For $d\ge 3$, such a path includes at least $4$ vertices. Let's put these $4$ vertices (denoted as vertices $1$, $2$, $3$, and $4$) in different layers, as shown in Fig.~\ref{fig:innerlayer3}(a). The $2$-path rule requires vertices $1$ and $3$ to have at least one more $2$ path between them, and the same for vertices $2$ and $4$, so the minimal graph 
becomes Fig.~\ref{fig:innerlayer3}(b) with two additional vertices $5$ and $6$. Now assume that one of the two inner layers has only two vertices; take it to be the left inner layer in Fig.~\ref{fig:innerlayer3}(b). Then we must have $5\sim6$ for vertices $1$ and $6$ to satisfy the $2$-path rule. Then the no $1221$ rule is effective, so there should be at least one more vertex $7$ in the right inner layer which hosts one more $2$-path. Take without loss of generality this path to be $2\sim7\sim4$, the minimal graph now becomes Fig.~\ref{fig:innerlayer3}(c). Then we must have $5\sim7$ for vertices $1$ and $7$ to satisfy the $2$-path rule, so the minimal graph becomes Fig.~\ref{fig:innerlayer3}(d). But then the six vertices $2,3,4,5,6,7$ form a $K_{3,3}$ subgraph, so the no $K_{3,3}$ rule is violated. Therefore, the assumption that one of the two inner layers has only $2$ vertices does not hold, and so any inner layer must contain at least $3$ vertices.

As for the two outer layers, 
each of them has at least $1$ vertex. So a graph with a diameter $d\ge 3$ must contain at least $3(d-1)+2=3d-1$ vertices. Thus, we arrive at a constraint between $d$ and $N$:
\begin{align}\label{eq:N-d}
N\ge 3d-1.
\end{align}
In particular, a graph with $d=3$ must have at least $8$ vertices, and graphs with $8$, $9$ or $10$ vertices must have $d\le 3$.

\subsection{Minimal subgraphs}

\begin{figure}[!htb]
(a)~ \scalebox{0.36}[0.36]{\includegraphics{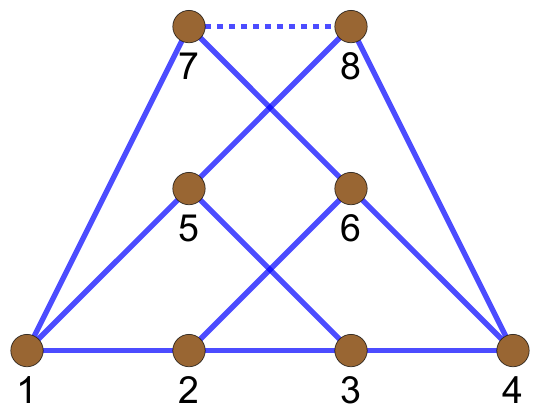}} ~
(b)~ \scalebox{0.36}[0.36]{\includegraphics{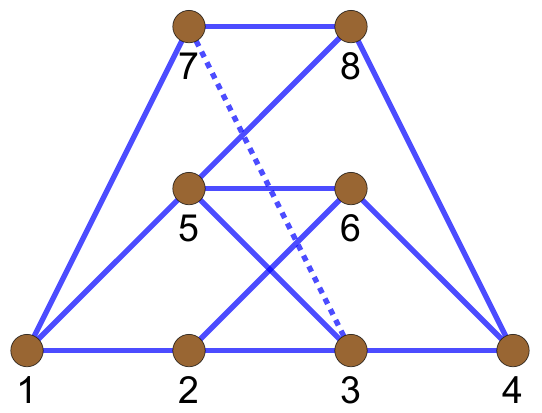}}
\caption{
The minimal subgraphs with $d\ge 3$: (a) with no 1221 subgraphs; (b) with at least one $1221$ subgraph.}
\label{fig:minimal}
\end{figure}

Based on the analysis above, we now construct the ``minimal subgraphs'', namely, the subgraphs that each integrable graph must contain, for $d\ge 3$. They will be useful when generating candidate graphs in the next section. We are going to consider separately two cases according to the property that if the graph is $1221$-free or not. The minimal subgraph in each case is shown in Fig.~\ref{fig:minimal}. Each graph contains $8$ vertices; if two vertices may or may not have an edge, they are connected by a dashed line. Except the drawn edges, no other edges are possible among the $8$ vertices (so these minimal subgraphs are actually induced subgraphs).

We now describe how the two minimal subgraphs are obtained. For each case, we can start from the graph of $6$ vertices, namely, the graph Fig.~\ref{fig:innerlayer3}(b). For the $1221$-free case, we have $5\nsim6$, so by the $2$-path rule there must exist vertices $7$ and $8$ such that $1\sim7\sim 6$ and $4\sim 8\sim5$. Then for the graph to be $1221$-free, we must have $2\nsim8$ and  $3\nsim7$. The vertices $7$ and $8$ may or may not have an edge. We thus arrive at Fig.~\ref{fig:minimal}(a). For the case with $1221$, without loss of generality we choose this $1221$ to be among the vertices $1$ to $6$, so $5\sim 6$. Now by the no $1221$ rule, we must add one vertex to form a new $2$-path, and without loss of generality we take this path as $4\sim 8\sim 5$. Now the no $K_{33}$ rule gives $2\nsim 8$, and then the $2$-path rule for vertices $1$ and $8$ requires $1\sim 7\sim 8$ with another vertex $7$ added. It remains to determine whether there are edges between $3$ and $7$ and between $6$ and $7$. By the no $K_{33}$ rule it is clear that both edges cannot exist simultaneously, since otherwise no $K_{33}$ rule would be broken. The vertices $3$ and $6$ are equivalent, so we can assume without loss of generality that $6\nsim7$. The vertices $3$ and $7$ may or may not have an edge. We then arrive at Fig.~\ref{fig:minimal}(b).

\section{A search for integrable graphs with $N\le 11$}

We now employ the graph analysis from the previous section to formulate an algorithm for searching candidate graphs (recall that a ``candidate graph'' means a graph that satisfies all the four rules stated in Sec. IIC) at a given vertex number $N$. Applying this algorithm, we identify all candidate graphs with $N\le 13$. We further perform analysis on those candidate graphs with $N\le 11$.

\subsection{Algorithm on enumeration of candidate graphs}

As $N$ increases, the number of possibilities of graphs increase rapidly. We therefore formulate an algorithm on a search for candidate graphs suitable for implementation on a computing program. Such an algorithm is based on the minimal subgraphs described in Sec. IIIB.

Here we describe this algorithm. 
Note that since candidate graphs of diameters $d=1$ and $d=2$ are completely identified, we only need to consider candidate graphs with $d\ge 3$. All these graphs must contain one of the minimal subgraphs (with $8$ vertices) in Fig.~\ref{fig:minimal}. We then start from drawing one minimal subgraph (either $1221$-free or not). Then we add $N-8$ vertices to the graph, and enumerate all possibilities that an edge may be added among them. After a graph is generated, we test whether the graph is connected, bipartite, and obeys all the four rules. In particular, for the $1221$-free case, we only need to test the $2$-path rule and the absence of an $1221$ subgraph, since the no $K_{3,3}$ rule and the no $1221$ rule are automatically satisfied for a $1221$-free graph. Whereas for the case with $1221$, we need to test the $2$-path rule, the no $K_{3,3}$ rule and the no $1221$ rule. If a graph passes all these tests, we compare it to all the existing graphs in the list of candidate graphs, and add it to the list if it is not isomorphic to any existing graphs.

The above algorithm can be conveniently performed by software equipped with a graph manipulation toolkit, e.g. Mathematica \cite{Mathematica}. The number of undetermined edges is $E=1+8(N-8)+(N-8)(N-9)/2=N(N-1)/2-27$ for the $1221$-free case or for the case with $1221$. The total number of cases to be searched is therefore $2^E$. Compared to a brute-force search of graphs with no restrictions whose number of possible edges is $N(N-1)/2$, this algorithm increases the efficiency by a factor $2^{27}$. For $N=11$, $E=28$, and the running time of the search algorithm is affordable; for $N=12$, $E=39$, and the running time becomes prohibitively long for practical computation. We then adapted the algorithm as below. First, we fix the diameter $d$ (for example, for $N=8$, the sequence can only be $1331$; for $N=9$, the sequence can be $1341$ or $1332$; For $N=10$, the sequence can be $1351$, $1441$, $1432$, $1342$, $1333$, or $2332$ \cite{note-inversion}). Second, we enumerate all possible sequences of vertex numbers in each layer $N_0 N_1 \ldots N_d$. Finally, we consider edges between adjacent layers at each sequence (again, the edges of the minimal graphs are fixed in advance). In this way only bipartite graphs are enumerated before tests of the rules, thus increasing the efficiency further. Using this improved algorithm, we achieved a complete enumeration of candidate graphs with up to $N=13$ vertices. For $N=14$, this algorithm also takes too long to finish.

The code for the algorithm described above, along with all the candidate graphs with $N\le 13$ generated by this algorithm, is openly available at \cite{github}. In Table \ref{table1}, we list the numbers of candidate graphs with and without a $1221$ subgraph from $N=8$ to $N=13$ and with $d\ge 3$. We see that the number of graphs with a $1221$ subgraph increases rapidly with $N$, whereas the number of $1221$-free graphs remains small up to $N=13$. For the three odd $N$'s, there are no $1221$-free graphs. For each of the three even $N$'s there is a single $1221$-free graph; in fact, they are just the graphs $(K_2)^3$, $K_2\times K_{2,3}$ and $K_2\times K_{2,4}$ which are known to be integrable. In other words, for vertex number up to $N=13$, excluding the known integrable graphs, any candidate graph has a $1221$ subgraph. One may wonder whether this holds for all $N$; however, in Sec. V, we show that there indeed exist $1221$-free candidate graphs, beyond the known integrable ones, for larger $N$.


\begin{table}[]\label{}
\caption{Numbers of 
candidate graphs with and without $1221$ subgraph for vertex numbers $N=8$ to $13$ with diameters $d\ge3$. \label{table1}}
\begin{tabular}{c|c|c}
  \hline
   \hline
    Vertex number $N$  & Number of   &  Number of graphs \\
      &  $1221$-free graphs &   with $1221$\\
      \hline
      8 &  1 & 1 \\
       \hline
   9 & 0 & 2 \\
          \hline
   10 & 1 & 7 \\
          \hline
   11 & 0 & 7 \\
 \hline
    12 & 1 & 30 
    \\
 \hline
    13 & 0 & 46 
    \\
 \hline
 \hline
\end{tabular}
\end{table}

\subsection{Analysis on Graphs with $N= 10$ and $11$}

In this subsection we perform analysis on candidate graphs with $N\le 11$.

As stated before, all $d=1$ and $2$ candidate graphs are completely determined ($K_2$ and $K_{2,N-2}$), and these graphs are indeed integrable. So below we focus on graphs with $d\ge 3$. For $d=3$, according to \eqref{eq:N-d}, we must have $N\ge 8$. For $N=8$ and $9$, as listed in Table \ref{table1}, there are $1$ and $2$ candidate graphs, respectively. The single $N=8$ candidate graph is the cube graph $(K_{2})^3$ which is indeed integrable; it has been studied in detail in \cite{MTLZ-2020}. The two candidate graphs with $N=9$ were analyzed in \cite{nogo-2022}; it turns out that they are not integrable. Below we analyze candidate graphs with $N=10$ and $11$.

\begin{figure*}[!htb]
(a)~ \scalebox{0.28}[0.28]{\includegraphics{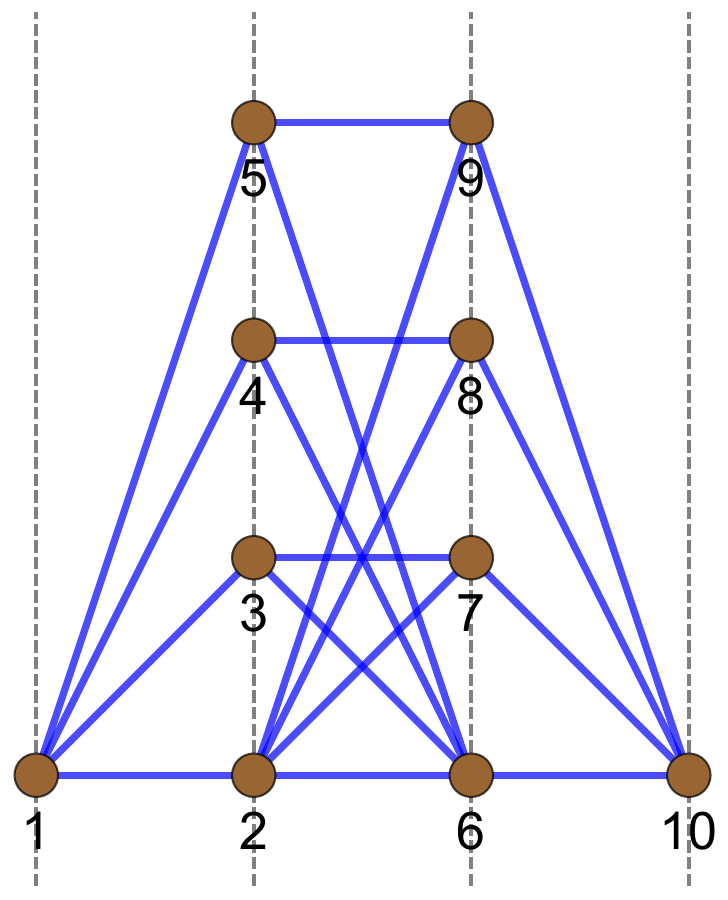}}
(b)~ \scalebox{0.28}[0.28]{\includegraphics{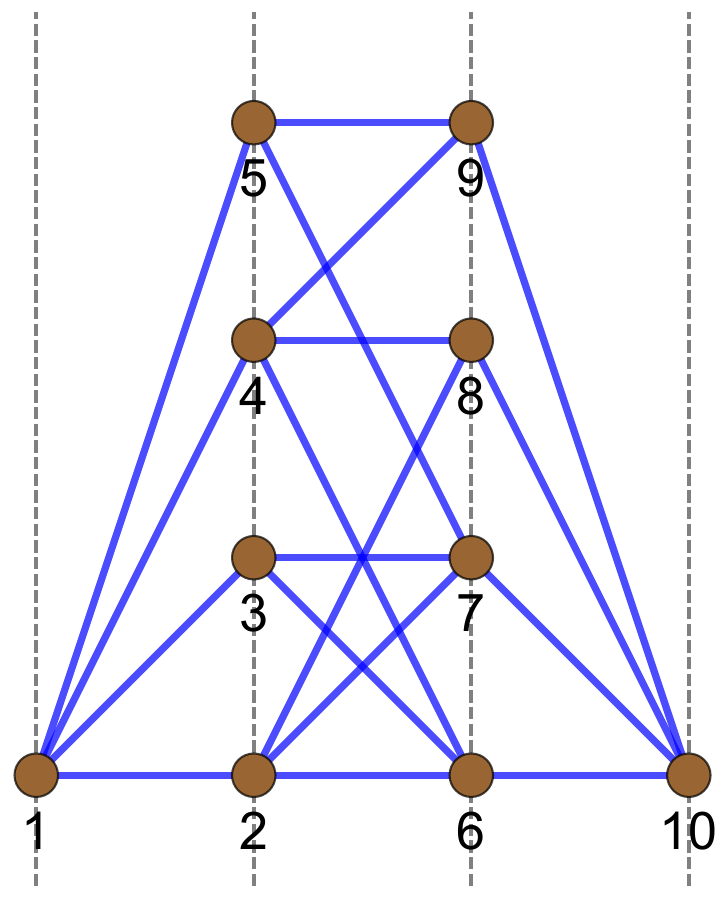}}
(c)~ \scalebox{0.28}[0.28]{\includegraphics{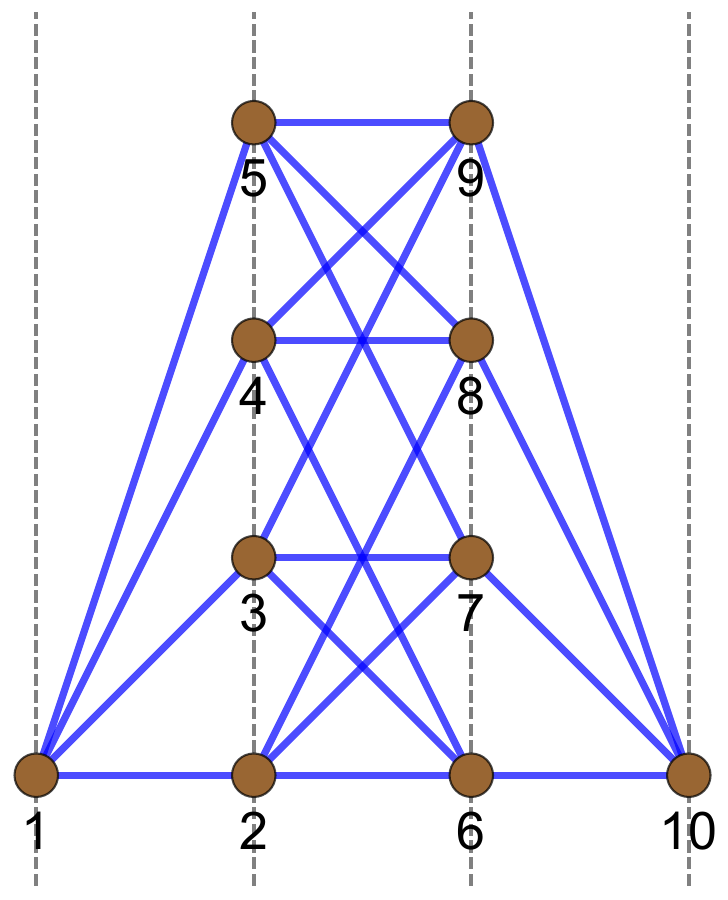}}
(d)~ \scalebox{0.28}[0.28]{\includegraphics{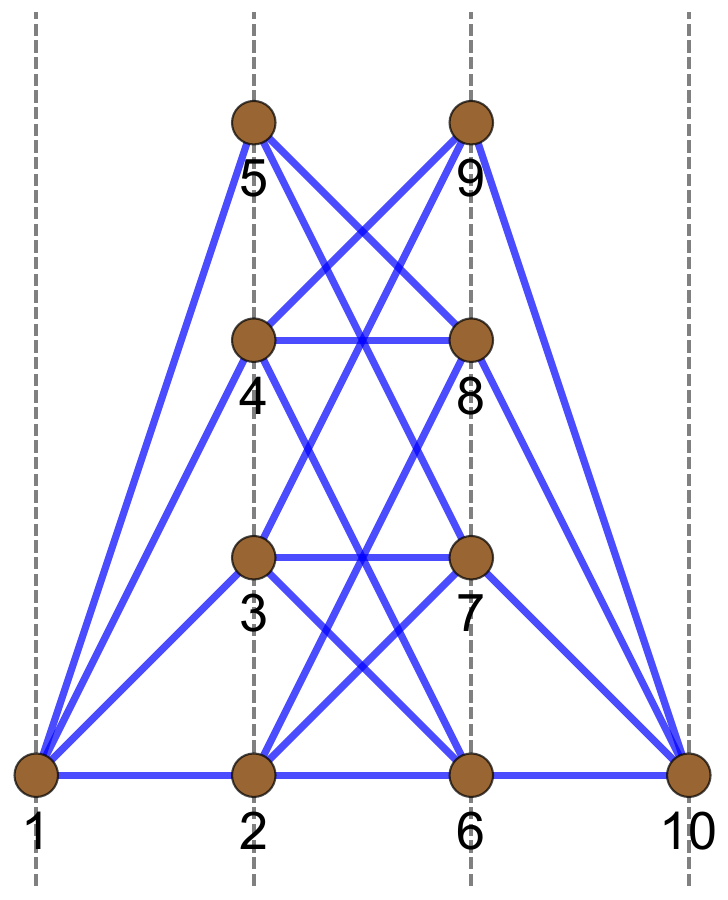}}
(e)~ \scalebox{0.28}[0.28]{\includegraphics{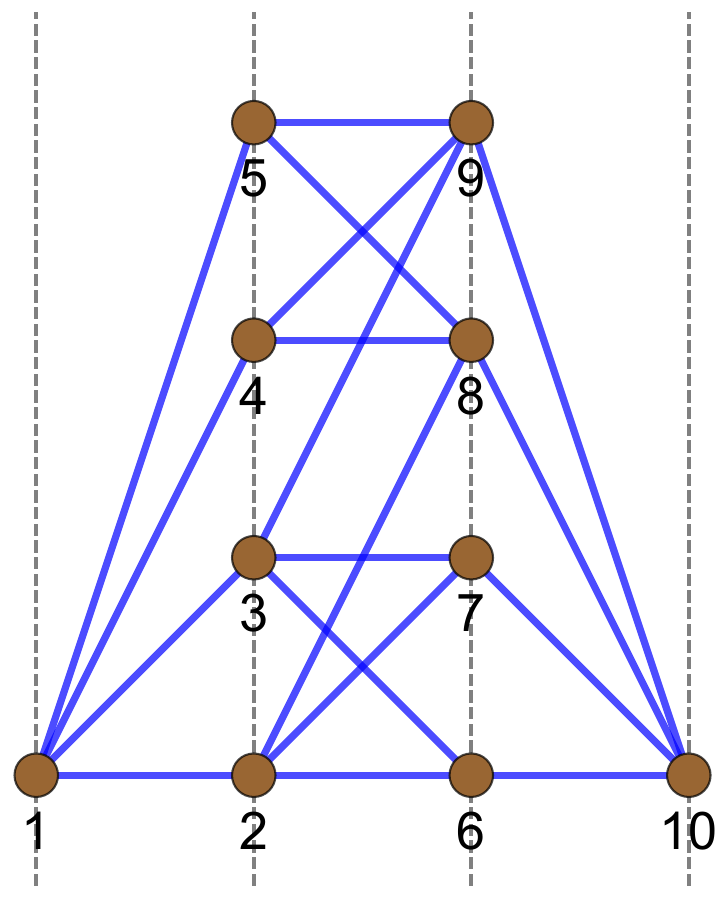}}
(f)~ \scalebox{0.28}[0.28]{\includegraphics{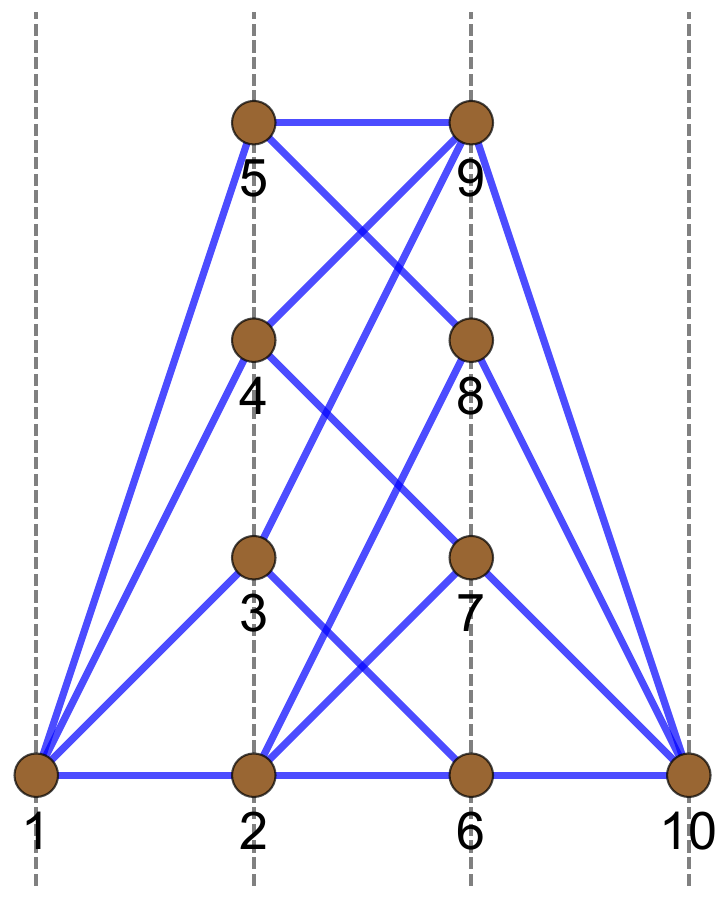}}
(g)~ \scalebox{0.28}[0.28]{\includegraphics{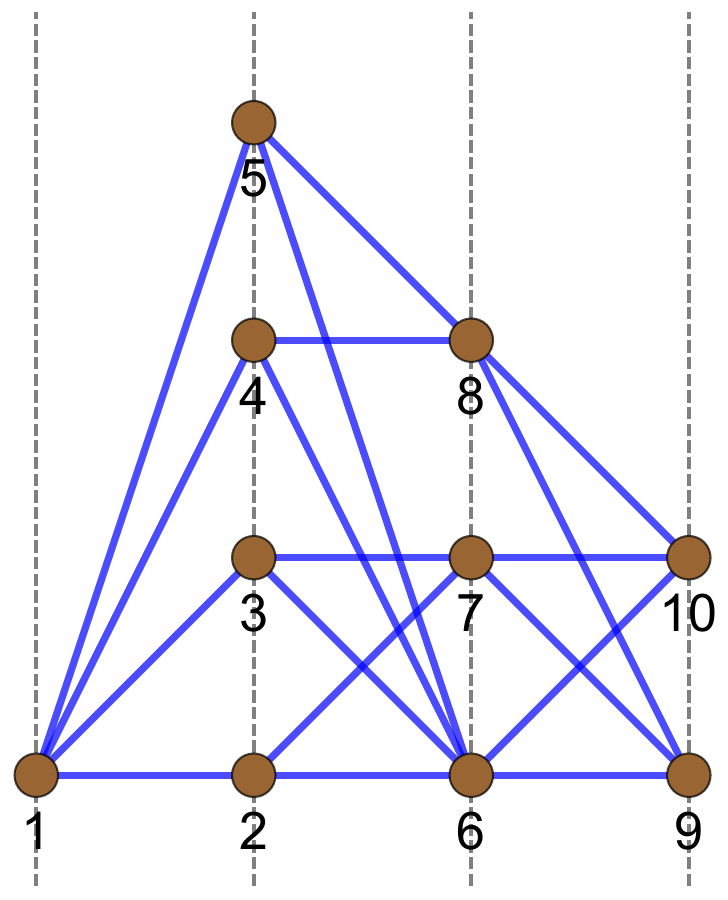}}
(h)~ \scalebox{0.35}[0.35]{\includegraphics{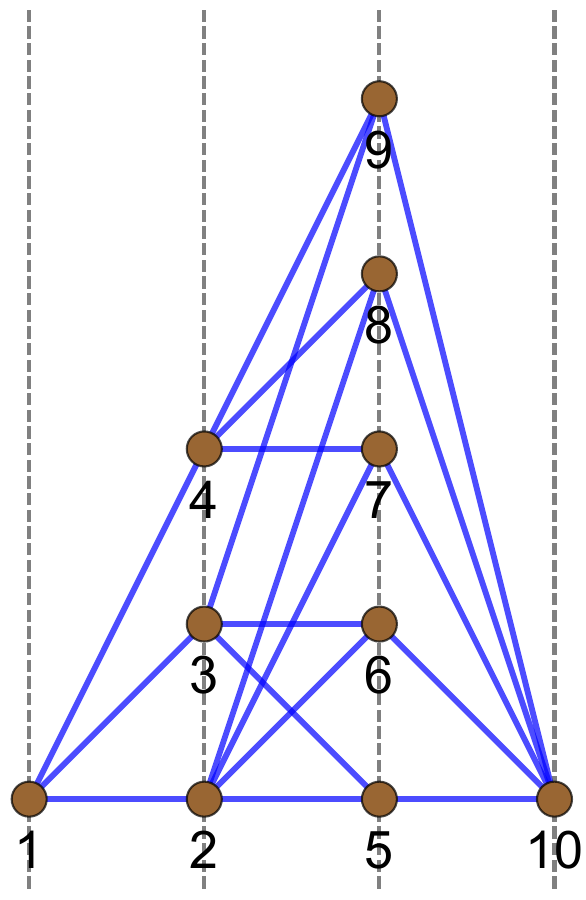}}
\caption{Candidate graphs 
with $N=10$ and $d=3$: (a) $1441$-$1$, (b) $1441$-$2$, (c) $1441$-$3$, (d) $1441$-$4$, (e) $1441$-$5$, (f) $1441$-$6$, (g) $1432$
, (h) $1351$. }
\label{fig:N=10-d=3}
\end{figure*}

The $8$ candidate graphs with $N=10$ and $d=3$ are drawn in Fig.~\ref{fig:N=10-d=3}. In Table \ref{table2} we list two types of data for each of these graphs: the total number of edges, and the list of degree of each vertex (namely, number of edges at the vertex). They have number of edges from $17$ to $20$. The graph with $17$ edges, Fig.~\ref{fig:N=10-d=3}(f), is $1221$-free. This graph is guaranteed to be integrable, since it is $K_2\times K_{2,3}$, namely, the Cartesian product of two integrable graphs $K_2$ and $K_{2,3}$. Other $7$ graphs all have $1221$ subgraphs.

\begin{table}[]\label{}
\caption{The total number of edges and the list of degree of each vertex for each $N=10$ graph shown in Fig.~\ref{fig:N=10-d=3}. \label{table2} }
\begin{tabular}{c|c|c}
  \hline
   \hline
    Graph & Number of edges & Degree of each vertex\\
      \hline
      $1441$-$6$ & $17$ & $4,4,4,4,3,3,3,3,3,3$ \\
   $1351$ & $18$ & $5,5,4,4,3,3,3,3,3,3$ \\
   $1441$-$1$ & $18$ & $5,5,4,4,3,3,3,3,3,3$ \\
   $1441$-$2$ & $18$ &  $4,4,4,4,4,4,3,3,3,3$ \\
   $1441$-$5$ & $18$ & $4,4,4,4,4,4,3,3,3,3$ \\
   $1432$-$1$ & $18$ & $6,4,4,4,3,3,3,3,3,3$ \\
   $1441$-$4$ & $19$ & $4,4,4,4,4,4,4,4,3,3$ \\
   $1441$-$3$ & $20$ & $4,4,4,4,4,4,4,4,4,4$ \\
 \hline
 \hline
\end{tabular}
\end{table}

Recall that the four rules are only necessary conditions for a graph to be integrable; whether a candidate graph is indeed integrable depends on whether there are nontrivial solutions to the equations obtained from the two properties \eqref{cycle-bar-A} and \eqref{multipath-bar-A}. In particular, one can follow the procedures described near the end of Sec. IIB. Here, we performed analysis on each of the $7$ graphs with $1221$ for $N=10$; in the next subsection, we present details of analysis on one of these graphs, namely Fig.~\ref{fig:N=10-d=3}(b), as an example. 
We find that some graphs do not allow a consistent set of the $r$ factors, so they are not integrable and no further analysis on them are needed. Some other graphs have possible sets of $r$ factors, e.g. the graph Fig.~\ref{fig:N=10-d=3}(b). For these graphs, the equations on $\gamma^{ab}$ are then considered. These equations are too complicated to handle analytically, so we solve them numerically (see next subsection for details). In each graph, only trivial solutions, with some of the $\gamma^{ab}$ being zero, are found. (As a test, we also performed numerical solving procedure of the equations on $\gamma^{ab}$ of the $K_2\times K_{2,3}$ graph which is known to be integrable, and nontrivial solutions are obtained.) This indicates that none of these graphs admits a solution, namely, the only integrable graph for $N=10$ is $K_2\times K_{2,3}$. We also performed analysis on each of the $7$ graphs with $1221$ for $N=11$. Again, we found no solutions for any of these graphs.

As shown in Table \ref{table1}, the numbers of candidate graphs for $N\ge 12$ increase rapidly, therefore a full analysis on graphs with $N=12$ or $13$ require tremendous efforts and is left for future. 

\subsection{Example: analysis on one $N=10$ graph}

\begin{figure*}[!htb]
\scalebox{0.5}[0.5]{\includegraphics{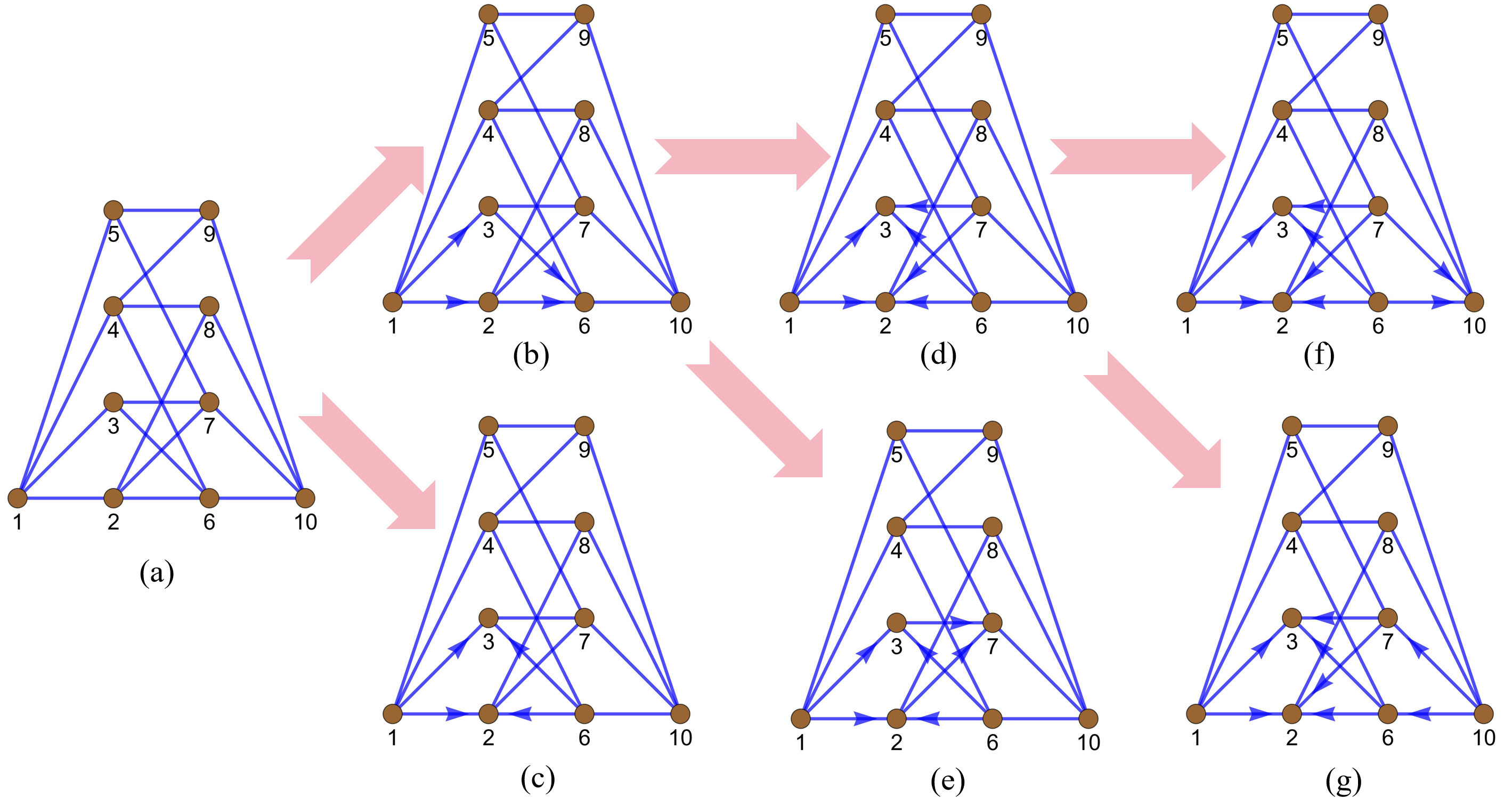}}
\caption{Illustration of the consideration of possible arrow configurations using a ``branch'' strategy. Starting from the undirected graph (a), arrows are added gradually to generate a tree of all possible graphs. Constraints of $r$ factors are tested for each case, and once a contradiction is reached, one immediately knows that the considered branch has no solutions of $r$ factors, and thus can turn to consider other branches.}
\label{fig:1441-2-branches}
\end{figure*}

In this subsection, as an illustrative example, we present details of analysis on one candidate graph with $N=10$, namely, the graph shown in Fig.~\ref{fig:N=10-d=3}(b) and also in Fig.~\ref{fig:1441-2-branches}(a). Analysis on other candidate graphs with $N=10$ and $11$ were performed similarly.

We start by considering possible solutions all $r$ factors defined in  \eqref{A-LO-transf-correspond-2}-\eqref{connect-bipart}. For each 4-cycle, there are two $r$ factors. Totally there are $36$ $r$ factors, i.e twice the total number of 4-cycles.

For the original undirected graph without specifying any arrow directions, there already exists a set of constraints on these $r$ factors. They come from the structures 
between pairs of vertices of distance $2$. For any such pairs in the graph in Fig.~\ref{fig:1441-2-branches}(a), there are either three or two $2$-paths. For example, consider $2$-paths between vertices in the first and third layers. Between vertices 1 and 6, there exist three $2$-paths ($1\sim 2\sim6$, $1\sim3\sim6$, and $1\sim4\sim6$). 
These paths form three distinct $4$-cycles, yielding three $r$ factors $r_{1263}$, $r_{1264}$ and $r_{1364}$ between vertices 1 and 6. Similarly, the three $2$-paths between vertices 1 and 7 correspond to three $r$ factors $r_{1273}$, $r_{1375}$ and $r_{1275}$. But only two paths link vertices 1 and 8, forming a single 4-cycle, thus there is only one $r$ factor $r_{1284}$ between vertices 1 and 8; the same goes for vertices 1 and 9. The $2$-paths between vertices in the second and fourth layers, and those between vertices in the same layer can be analyzed analogously.

For a pair of vertices of distance $2$ with two $2$-paths between them, the multipath property \eqref{multipath-gamma} fixes an $r$ factor to be $-1$. Taking vertices 1 and 8 as an example, and denote the sign factors between paths $1\sim2\sim8$ and $1\sim4\sim8$ as $r_{1284}$. The multipath condition \eqref{multipath-gamma} between vertices 1 and 8 reads:
\begin{equation}
\sqrt{|\gamma_{12}  \gamma_ {28}|}+r_{1284} \sqrt{| \gamma _{14}  \gamma _{48}|}=0.
\end{equation}
Recall that the $r$ factors are signs which can only take the values of $1$ or $-1$. Consequently, to satisfy this equation, $r_{1284}$ here must be $-1$. Similarly, we have:
\begin{equation}\label{eq:r2}
\begin{split}
& r_{1284} = r_{1495}= r_{3607} = r_{5709} = r_{2157} =  r_{3146}= r_{3157} \\
 & = r_{4159} = r_{6490} = r_{7280} = r_{7590} = r_{8490} =-1,
\end{split}
\end{equation}
where for notation simplicity in the subscript of $r$ we wrote an index 0 for vertex 10.

For a pair of vertices of distance $2$ with three $2$-paths between them, the multipath property \eqref{multipath-gamma} contains three terms.
For example, the multipath condition for vertices 1 and 6 reads:
\begin{equation}
\sqrt{|\gamma_{12}   \gamma_{26}|}+r_{1263}   \sqrt{|\gamma_{13}   \gamma_{36}|}+r_{1264} \sqrt{|\gamma_{14}\gamma_{46}|}=0,
\end{equation}
and $r_{1364}$ is related to $r_{1263}$ and $  r_{1264}$ as
\begin{equation}
r_{1364}= r_{1263} r_{1264}.
\end{equation}
From these equations, it is evident that among these three $r_{1263}$, $r_{1264}$ and $r_{1364}$ two must be $-1$, and one must be $1$. This is equivalent to the two equations of $r_{1263} + r_{1264} + r_{1364} = -1$ and $ r_{1263}   r_{1264}   r_{1364} = 1$, which is symmetric in the three $r$ factors. This analysis can be extended to other cases involving two vertices connected by three paths, giving $16$ equations as follows:
\begin{equation}\label{eq:r3}
\begin{aligned}
&r_{1263} + r_{1264} + r_{1364} = -1, & r_{1263}   r_{1264}   r_{1364} = 1, \\
&r_{1273} + r_{1375} + r_{1275} = -1, & r_{1273}   r_{1375}   r_{1275} = 1, \\
&r_{2607} + r_{2608} + r_{2708} = -1, & r_{2607}   r_{2608}   r_{2708} = 1, \\
&r_{4608} + r_{4609} + r_{4809} = -1, & r_{4608}   r_{4609}   r_{4809} = 1, \\
&r_{2136} + r_{2137} + r_{2637} = -1, & r_{2136}   r_{2137}   r_{2637} = 1, \\
&r_{2148} + r_{2146} + r_{2648} = -1, & r_{2148}   r_{2146}   r_{2648} = 1, \\
&r_{6270} + r_{6273} + r_{6370} = -1, & r_{6270}   r_{6273}   r_{6370} = 1, \\
&r_{6280} + r_{6480} + r_{6284} = -1, & r_{6280}   r_{6480}   r_{6284} = 1.
\end{aligned}
\end{equation}
The reason to choose this symmetric way to present the conditions is that it is easier to be implemented in a computing program. Note that in the constraints \eqref{eq:r2} and \eqref{eq:r3}, all the $36$ $r$ factors appear.

The above constraints \eqref{eq:r2} and \eqref{eq:r3} on the $r$ factors does not depend on the directions of edges. We now start to specify the directions, namely, putting in arrows in the original undirected graph, like in Fig.~\ref{fig:4-loop-graph}. This gives a new set of constraints on the $r$ factors; specifically, depending on whether the orientation of a 4-cycle is non-bipartite or bipartite (see Fig.~\ref{fig:4-loop-graph}), the two $r$ factors have the same or opposite signs, respectively. Recall that only those configurations of arrows whose constraints are all consistent (namely, without any contradictions) can possibly be integrable. To find all possible solutions of the $r$ factors, the most direct way is to enumerate possible configurations of all arrows in the graph, and determine if there are solutions of $r$ factors in each case. Since the number of possible configurations is usually very large, this strategy would be time-consuming. We instead adopt a more efficient strategy, which we call the ``branch'' strategy: starting from the undirected graph we add arrows progressively, so the constraints are also added progressively. At each step, all possible ways of adding arrows should be considered. Then a tree of graphs with more and more branches will be generated gradually, with its root being the undirected graph and with each branch corresponding to different configurations of the added arrows. If a contradiction arises for a particular arrow configuration at a step of adding arrows, we can conclude that any directed graph containing this arrow configuration is not integrable and proceed to other branches. If, however, arrows have been added to the entire graph without any contradiction arising, then this arrow configuration yields a solvable set of $r$ factors. After considering all branches (namely, all possible ways of adding arrows), this strategy finally generates a tree with each leaf of the tree either having contradictions or containing a solvable set of $r$ factors.



Fig.~\ref{fig:1441-2-branches} illustrates this ``branch'' strategy in detail. In the undirected graph Fig.~\ref{fig:1441-2-branches}(a), we first add arrows starting from the $1\sim2\sim6\sim3\sim 1$ cycle, which may have a bipartite orientation or a non-bipartite orientation, as shown in Fig.~\ref{fig:1441-2-branches}(b) and (c), respectively
. For Fig.~\ref{fig:1441-2-branches}(b), from Eq.~\eqref{connect-bipart}, we get the additional relation
\begin{align}
&r_{1263}=-r_{2136}.
\end{align}
For Fig. \ref{fig:1441-2-branches}(c), from Eq.~\eqref{A-LO-determinants-tilde}, we get instead
\begin{align}
&r_{1263}=-r_{2136}.
\end{align}
We then add arrows in the edges $2\sim7$ and $3\sim 7$ from each of Fig.~\ref{fig:1441-2-branches}(b) and (c). For example, in Fig.~\ref{fig:1441-2-branches}(b), there are two possible ways of arrow configurations, shown in  Fig.~\ref{fig:1441-2-branches}(d) and (e).  
In Fig.~\ref{fig:1441-2-branches}(d), we get additional constraints
\begin{equation}
\begin{aligned}
& r_{1273}=-r_{2137}, \quad r_{2637}=-r_{6273},
\end{aligned}
\end{equation}
whereas in Fig.~\ref{fig:1441-2-branches}(e), we get
\begin{equation}
\begin{aligned}
& r_{1273}=r_{2137}, \quad r_{2637}=r_{6273}.
\end{aligned}
\end{equation}
We then add arrows on the edges $6\sim10$ and $7\sim 10$ from all possible cases. For example, in Fig.~\ref{fig:1441-2-branches}(d), there are two possible ways of arrow configurations, shown in Fig.~\ref{fig:1441-2-branches}(f) and (g), each yielding additional constraints on the $r$ factors.

All the graphs shown in Fig.~\ref{fig:1441-2-branches} has no contradictions of the equations of $r$ factors, but when adding more arrows it's possible that a contradiction is reached. Continuing this process of adding arrows until all branches are considered, we can identify all possible cases that host solvable sets of $r$ factors.  

\begin{figure}[htbp]
	\centering
	\begin{subfigure}{0.22\textwidth}
		\includegraphics[width=\linewidth]{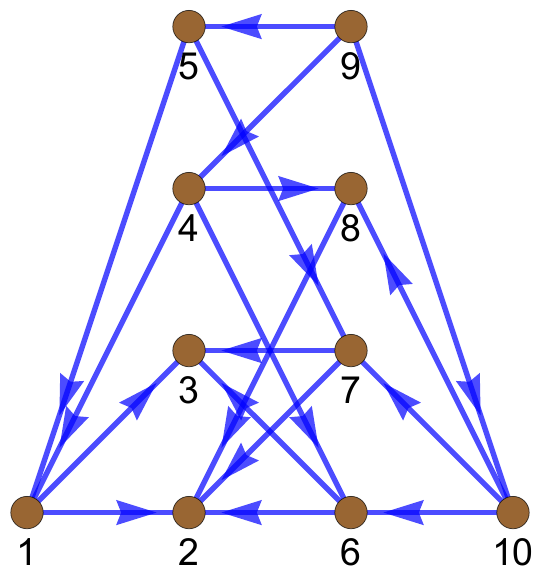}
		\caption{ }
		\label{fig:bipartite}
	\end{subfigure}
	\hspace{3.0cm}
	\begin{subfigure}{0.22\textwidth}
		\includegraphics[width=\linewidth]{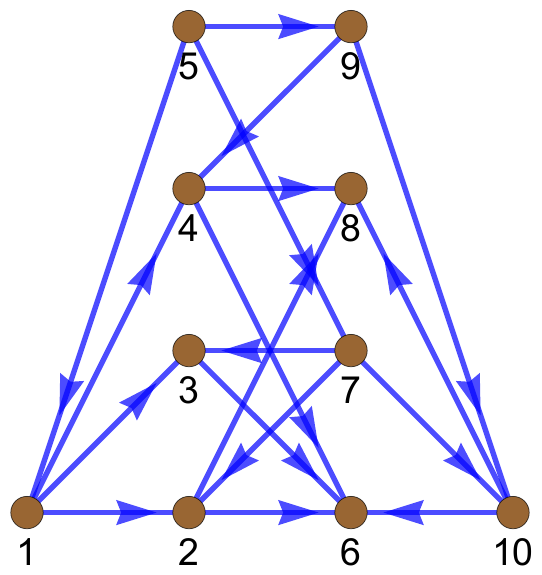}
		\caption{ }
	\end{subfigure}
	\caption{The two directed graphs for which the $r$ factors have solutions.}
	\label{fig:possiblegraph}
\end{figure}
It turns out that only two directed graphs yield solvable $r$ factors; they are depicted in Fig.~\ref{fig:possiblegraph}. For each graph there is more than one set of solutions of $r$ factors. For example, in Fig.~\ref{fig:possiblegraph}(a),
there are two sets of solutions for the $r$ factors. One set has $r_{1263} =r_{1273}=r_{2607}= r_{4608}= r_{2637}= r_{2648}=r_{6270}= r_{6284}=1$, whereas all other $r$ factors are $-1$. Another set has $r_{1264}= r_{1273}=r_{2608}= r_{4608}=r_{2136}=r_{2146}= r_{6273} =r_{6280}=1 $, whereas all other $r$ factors are $-1$.

For the two directed graphs with the solutions of $r$ factors, we then checked if there is a solution to the system of equations of $\sqrt{|\gamma^{ab}|}$ for all pairs of vertices at distance $2$. The number of these pairs is 20, so there are 20 equations of $\sqrt{|\gamma^{ab}|}$, each being of the form \eqref{multipath-gamma}. Analytical solution of such a large set of equations is difficult, so we instead looked for numerical solutions. Specifically, in Mathematica we write out all equations and specifies the initial values of all $\sqrt{|\gamma^{ab}|}$, and use the ``Findroot'' function to solve the system numerically. After trying a large number of starting values, we find that no solution set of the $r$ factors in the two directed graphs have nontrivial solutions with all $|\gamma^{ab}|$ positive. Numerical solutions with some of $|\gamma^{ab}|$ being vanishingly small are found, which implies that these $\gamma^{ab}$ are zero, but these solutions are not allowed since by definition $|\gamma^{ab}|$ must be positive (there should be nonzero couplings on each edge of the graph). This indicates that the graph Fig.~\ref{fig:N=10-d=3}(b) is not integrable.


\section{Larger Graphs---$(0,2)$ graphs and their descendants}

Our analysis in the previous section indicates that the conjecture stated in Sec. I---namely, for MTLZ models the only integrable graphs are $K_2$, $K_{2,n\ge 3}$ and any Cartesian products constructed by them---holds for $N\le 11$. (Note that this conjecture can also be stated as: there exist only two types of ``fundamental integrable graphs'' $K_2$ (the two-state LZ model) and $K_{2,n\ge 3}$ (the fan models), and all integrable graphs can be constructed from these two by Cartesian products.) Does the conjecture remain true for any larger $N$
? As mentioned before, a general proof to eliminate all other graphs seems difficult to reach, whereas to disprove the conjecture, one has to look for a counterexample at a still larger $N$.

As $N$ increases, analysis faces a two-fold complexity. First, identifying all candidate graphs becomes time-consuming due to the rapidly increasing number of possible edges: our search algorithm involves $2^E$ possibilities where $E$ (the number of undetermined edges) is $O(N^2)$, despite being $2^{27}$ times more efficient than a brute-force search (as mentioned before, our algorithm identified all candidate graphs at $N=13$, but for $N=14$ such a complete search takes too long; of course, at $N\ge 14$ the algorithm can still generate candidate graphs by exploring a part of (instead of all of) the phase space of undetermined edges, for example, simply by randomly adding undetermined edges). Second, a  complete analysis of a given candidate graph with large $N$ also becomes difficult due to the graph's intrinsic complexity. 

Therefore, in this section, instead of a complete search, we take another approach to 
analyze larger graphs. 
The idea is to consider first a yet stricter set of rules. 

\subsection{$(0,2)$-graphs}

We consider a class of graphs satisfying a stricter rule---namely, candidate graphs that are also $K_{2,3}$-free. Note that forbidden $K_{2,3}$ means that any $K_{m,n}$ graphs with $m\ge 2$ and $n\ge 3$ and the $1221$ graph are also forbidden as subgraphs, since $K_{m\ge 2,n \ge 3}$ or the $1221$ graph contains $K_{2,3}$. So the no $K_{3,3}$ rule and the no $1221$ rule are automatically satisfied. This new ``no $K_{2,3}$ rule'' together with the $2$-path rule requires that any two vertices have either $0$ or $2$ common neighbours. $K_2$ and Cartesian products of any number of $K_2$, namely the hypercube graphs $Q_D\equiv(K_2)^{ D}$ ($D\ge 1$), indeed belong to this class, and are integrable. If $Q_D$ were the only graphs in this class, no more analysis on this class will be needed; it is tempting to think that this is true, because it appears difficult to construct by hand a candidate graph besides $Q_D$ that is $K_{2,3}$-free.

Somewhat surprisingly, it turns out that there do exist other graphs besides $Q_D$ that belong to this class of graphs with any two vertices having either $0$ or $2$ common neighbours. Such nontrivial cases exist only for large numbers of vertices $N$---the smallest such graph has $N=14$. Such a class of graphs was introduced by Mulder in \cite{Mulder-1979} and was named the {\it $(0,2)$-graph}; since then these graphs have been studied in the area of discrete mathematics \cite{Mulder-1979,Berrachedia-1999,Brouwer-2006,Brouwer-2009}. In \cite{Mulder-1979} $(0,2)$-graphs are proved to be regular (namely the degree of each vertex is the same; this degree is often called a valency). Brouwer classified all $(0,2)$-graphs with valencies at most $7$ in \cite{Brouwer-2006}, and with valency $8$ in \cite{Brouwer-2009}. The numbers of nonisomorphic bipartite $(0,2)$-graphs from valency $k=1$ up to valency $k=8$ are: $1,1,1,2,4,13,40,104$ \cite{Brouwer-2009}. At each valency $k$, there is the graph $Q_k$, so bipartite $(0,2)$-graphs other than $Q_k$ starts to exist at $k=4$, and its numbers from $k=4$ to $k=8$ are $1,3,12,39,103$. In particular, at $k=4$, there is a single graph except $Q_4$ with $N=14$. This graph is shown in Fig.~\ref{fig:(0,2)-14}(a), which we call the $1463$ graph (this name comes from the sequence of vertex numbers in each layer viewed from bottom to top). The structure of this $1463$ graph is such that the three bottom layers are the same as that of the $4$-dimensional hypercube $Q_4$ (which has a sequence $14641$), but in the top two layers two vertices are deleted, and some edges are rearranged in a way to obey the constraint of $(0,2)$-graphs.

A closer examination at the $1463$ graph in Fig.~\ref{fig:(0,2)-14}(a) shows that it is not integrable. Consider a $4$-cycle $1\sim 2\sim 3\sim4\sim1$ in this graph, the multipath property \eqref{multipath-gamma} yields two equations, each containing only two terms:
\begin{eqnarray}
&\sqrt{|\gamma^{12}\gamma^{23}|}+ r_{1234}  \sqrt{|\gamma^{14}\gamma^{34}|} = 0, \nn\\
&\sqrt{|\gamma^{12}\gamma^{14}|}+ r_{2341}  \sqrt{|\gamma^{23}\gamma^{34}|} =0. \label{}
\end{eqnarray}
These two equations can be satisfied only if the two $r$ factors are $ r_{1234} = r_{2341}  =-1$. This is possible for the non-bipartite orientation but not for the bipartite orientation. Thus, every $4$-cycle in the $1463$ graph (actually in any $(0,2)$-graph) must be of the non-bipartite orientation for the set of equations on $\gamma^{ab}$ to have solutions. It is not possible to put arrows in the graph to satisfy this; for example, if we put arrows on all edges from bottom to top in \ref{fig:(0,2)-14}(a), then the $4$-cycles that span three layers are of the non-bipartite orientation, but there are also $4$-cycles that span two layers (they exist in the upper two layers) which are of the bipartite orientation.

The above argument actually applies for any $(0,2)$-graph excluding hypercubes. In \cite{Berrachedia-1999} (see Lemma $6$ there) it was proved that any $(0,2)$-graph whose every $4$-cycle spans exactly $3$ layers is a hypercube, so any other $(0,2)$-graphs must have at least one $4$-cycle that spans $2$ layers. This means that when drawing a directed graph, there must always be $4$-cycles of the bipartite orientation. Therefore, except for hypercubes, any $(0,2)$-graph is not integrable.


\begin{figure}[!htb]
(a)~ \scalebox{0.5}[0.5]{\includegraphics{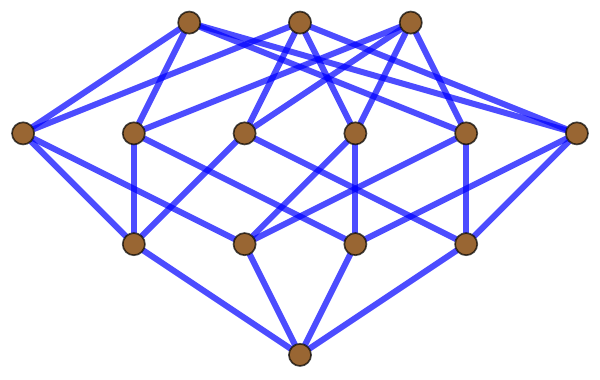}}
(b)~ \scalebox{0.5}[0.5]{\includegraphics{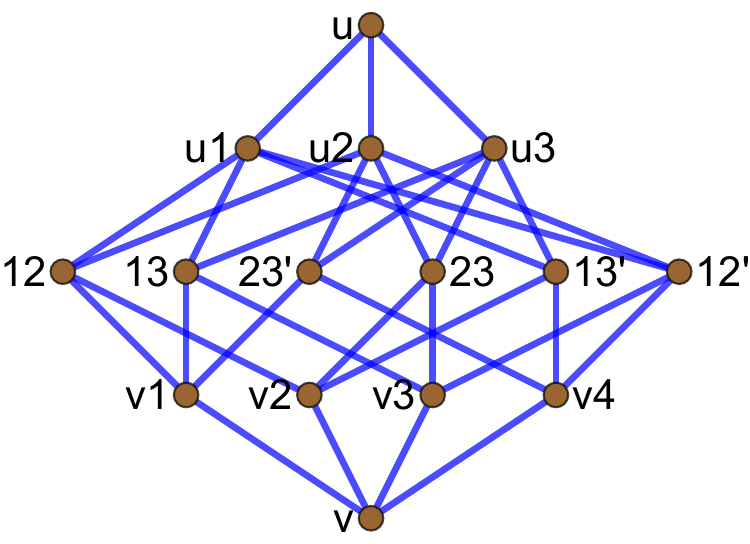}}
(c)~ \scalebox{0.5}[0.5]{\includegraphics{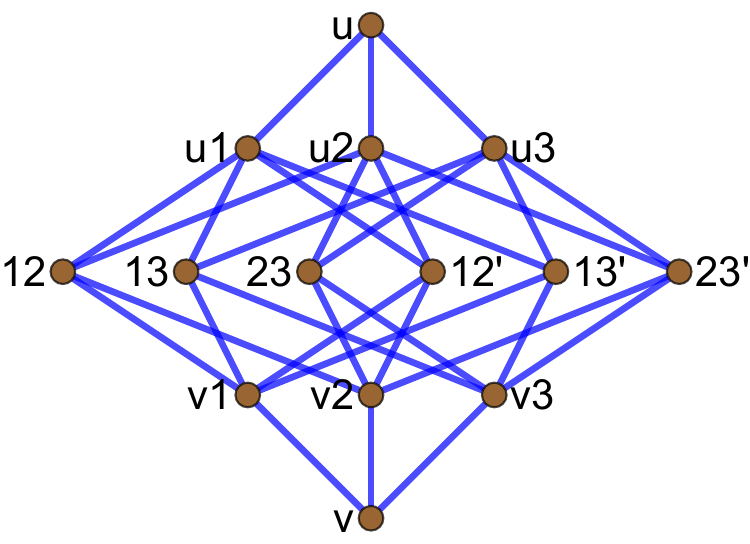}}
\caption{(a) The $1463$ graph, namely, the $(0,2)$-graph with $N=14$ and $k=4$. (b) The $14631$ graph: the graph in (a) with an additional vertex added at the top. (c) The $13631$ graph: the graph in (b) with one vertex deleted and some edges rearranged.}
\label{fig:(0,2)-14}
\end{figure}


\subsection{Descendants of $(0,2)$-graphs}

Although the $(0,2)$-graphs are not integrable, they provide insights on constructing candidate graphs which might be integrable. First, the argument above suggests that one may try to rescue those $4$-cycles of the bipartite orientation by adding or moving vertices while keeping the rules for candidate graphs still satisfied. Second, since the constraints of a $(0,2)$-graph is stricter than those of a candidate graph, a slight release of constraints of a $(0,2)$-graph possibly still leads to a candidate graph. Following these ideas, we consider two ``descendants'' of the $1463$ graph in this subsection. We will see that for these two graphs the conditions on the $r$ factors and on $\gamma^{ab}$ can be satisfied; this was not achieved by any candidate graph considered in Sec. IV with $N=10$ and $11$.

1. The $14631$ graph:


Consider a graph with $15$ vertices constructed in the following way. In the $1463$ graph in Fig.~\ref{fig:(0,2)-14}(a), we add an additional vertex on top of the three upmost vertices and connect it to the three upmost vertices. This results in Fig.~\ref{fig:(0,2)-14}(b) (the labels of vertices are named to observe symmetries of the graph). This graph now has $K_{2,3}$ subgraphs, but is still $1221$-free. We can also check that the $2$-path rule is still satisfied, so this graph is a candidate graph. Let's call it the $14631$ graph.

To proceed to analyze whether this graph is indeed integrable, let's consider a directed graph with all arrows from bottom to top in Fig.~\ref{fig:(0,2)-14}(b). Then all $4$-cycles that spans three layers of the graph are of the non-bipartite orientation, and all $4$-cycles that spans two layers of the graph are of the bipartite orientation. We next construct solutions of the $r$ factors and $\gamma^{ab}$. The set of constraints is complicated, so we seek simple solutions of $\gamma^{ab}$ by inspection. We find that the constraints on the $r$ factors and on $\gamma^{ab}$ can all be satisfied by a rather simple choice: setting $r_{u1,12,u2,12'} =r_{u1,13,u3,13'}=r_{u2,23,u3,23'}=1$, whereas all other $r$ factors are $-1$, and setting all $|\gamma^{ab}|$ to be the same as $|\gamma^{ab}|=\gamma$, except for those on the three edges at the vertex $u$ which are twice as large: $|\gamma^{u,u1}|=|\gamma^{u,u2}|=|\gamma^{u,u3}|=2\gamma$.

2. The $13631$ graph:

Now consider the graph in Fig.~\ref{fig:(0,2)-14}(c), which we call the $13631$ graph. This graph is constructed by removing the vertex $v4$ in Fig.~\ref{fig:(0,2)-14}(b) and rearrange some edges (note that positions of some vertices are interchanged so to make the figure appear more symmetric). This graph has $1221$ subgraphs, but one can verify that it remains a candidate graph.

Again, consider a directed graph with all arrows from bottom to top in Fig.~\ref{fig:(0,2)-14}(c). Similar to the $14631$ graph case, we find that the constraints on the $r$ factors and $\gamma^{ab}$ can all be satisfied by a simple choice of parameters: setting 
$r_{ui,ij,uj,ij'} =r_{vi,ij,vj,ij'}=r_{ui,ij,vi,ij'}=r_{vi,ij,ui,ij'}=1$ (where $i,j=1,2,3$ and $i\ne j$), whereas all other $r$ factors are $-1$, and setting all $|\gamma^{ab}|$ to be the same $|\gamma^{ab}|=\gamma$, except for those on the six edges incident on the vertex $u$ or $v$ which are twice as large: $|\gamma^{u,u1}|=|\gamma^{u,u2}|=|\gamma^{u,u3}|=|\gamma^{v,v1}|=|\gamma^{v,v2}|=|\gamma^{v,v3}|=2\gamma$.


For the two graphs above, if their sets of equations on $\bar A^{ab}$ with the given choices of $r$ factors have nontrivial solutions, then they would host MTLZ models. Unfortunately, for each of them this set of equations is too complicated to be handled analytically. We tried to solve the set numerically, and we did find solutions of $\bar{A}^{ab}$ when fixing to the equations on certain subgraphs, but when extending to the whole graph, the number of equations is too large and the numerical evaluation cannot be done. Besides, we do not identify simple choices of parameters to satisfy this set. (As a comparison, all currently known integrable graphs can be decomposed into fundamental integrable graphs $K_2$ and $K_{2,n\ge 3}$ in the Cartesian product sense, with Fig.~\ref{fig:N=10-d=3}(f) being an example, and for such an integrable graph one can utilize this decomposition to construct solutions of $\bar{A}^{ab}$ with simple choices of parameters.) This does not mean that the graphs are not integrable, since we only considered specific choices of arrow configurations and $r$ factors, and our numerical evaluation was not exhaustive. Future research is required to determine whether these two graphs are integrable or not. 


\section{Non-bipartite graphs}

We present a brief discussion on non-bipartite graphs which is omitted up to now. The reason for this omission is primarily that it is inconvenient to consider non-bipartite graphs within the layer graph scheme. In principle, non-bipartite graphs can be included in this scheme if we modify the definition of a layer graph by allowing edges connecting vertices in the same layer, but this results in a large growth of possibilities of edge configurations. Moreover, the currently known integrable graphs (hypercubes, fans, and their Cartesian products) are all bipartite. The following argument based on the rules of candidate graphs indicates that non-bipartite candidate graphs appear only at large vertex numbers. The no $K_3$ rule forbids a large family of non-bipartite graphs containing $3$-cycles, and the $2$-path rule dictates that $4$-cycles must flourish in any candidate graph, so a candidate graph containing $5$-cycles or other odd-cycles must be large. Therefore, integrability seems to favor bipartite graphs over non-bipartite graphs. Nevertheless, we do not find strict rules that forbid non-bipartite graphs to be integrable, and non-bipartite candidate graphs are definitely worth further research. Here, we identify a non-bipartite candidate graph at $N=16$. We believe that this is the smallest vertex number of non-bipartite candidate graphs, although a rigorous proof is lacking.

\begin{figure}[!htb]
\scalebox{0.5}[0.5]{\includegraphics{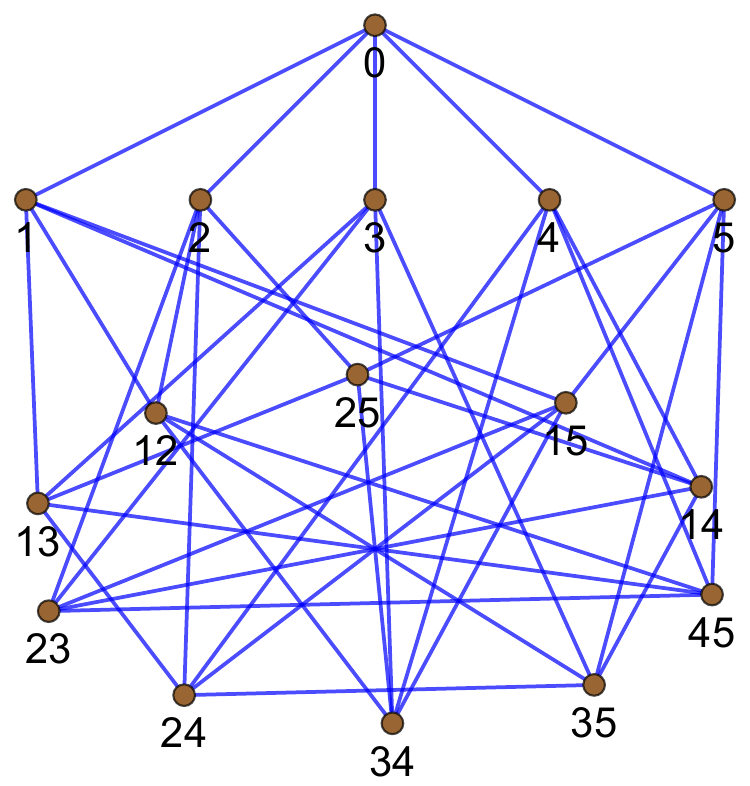}}
\caption{A non-bipartite $(0,2)$-graph with $N=16$, which is a candidate graph.}
\label{fig:1510}
\end{figure}

We again use knowledge of the $(0,2)$-graphs. In Fig.~\ref{fig:1510} the smallest non-bipartite $(0,2)$-graph which is $K_3$-free is shown. It is an $N=16$ graph with diameter $d=2$ and valency $k=5$ (it corresponds to the graph data on the fourth row in Table 2 of Ref. \cite{Brouwer-2006}). The graph looks cumbersome, but its structure can be better understood by noticing that there are three types of vertices labeled by $0$, $i$ ($i=1,2,3,4,5$), and $ij$ ($i,j=1,2,3,4,5$, $i\ne j$) in three layers from top to bottom, and vertices within each layer are symmetric. 
One can check that this graph has $5$-cycles (e.g. $0\sim 1\sim 13\sim 24 \sim 2 \sim 0$), so it is non-bipartite. As argued in Sec. VA, any $(0,2)$-graphs except for hypercubes are not integrable, and so is the graph in Fig.~\ref{fig:1510}. In future works it may be interesting to consider descendants of this graph like we did in Sec. VB.

Finally, we note that if non-bipartite integrable graphs do exist, their properties may be quite different from those of bipartite integrable graphs. For example, MTLZ models on bipartite integrable graphs must have symmetric transition probability matrices---a property required by bipartiteness \cite{cross-2017}; non-bipartite graphs, on the other hand, is not subject to this requirement, so their probability matrices may be asymmetric.


\section{Conclusions and discussions}

In this work, we present a systematic graph-theoretical search for integrable models in the MTLZ class. In this class, the hypercubes, the fans and their direct product models were known to be integrable, and there was the conjecture that no other integrable models exist. We formulate a search algorithm for graphs that may host MTLZ models, which is $2^{27}$ times more efficient than a brute-force search. Implementing this algorithm on Mathematica, we identify all candidate graphs for MTLZ models with vertex numbers $N\le 13$ (they are presented at \cite{github}). We further perform detailed analysis on the candidate graphs with $N=10$ and $11$. Our analysis supports the aforementioned conjecture for graphs with no more than $11$ vertices. For graphs with more vertices, we propose to consider descendants of ``$(0,2)$-graphs'', which may be promising candidates to violate the conjecture above. We expect our work to serve as a guideline to identify new exactly solvable MLZ models in the future.

Whether the conjecture is correct remains uncertain. It is tempting to view its validity for $N\le11$ as strong evidence that it holds for any $N$, but one cannot deny the possibility of exceptions for larger $N$ (for example, recall that $(0,2)$-graphs besides hypercubes appear only at $N\ge 14$). 
Hence, it is desirable to perform further studies on the candidate graphs with $N=12$ and $13$ found by our algorithm and on descendants of $(0,2)$-graphs. In particular, we expect descendants of $(0,2)$-graphs as promising candidates to break the conjecture, because they allow simple solutions of the $r$ factors and the $\gamma^{ab}$, a feature absent in any considered graphs with $10$ or $11$ vertices.

Researches on MTLZ models also call for new methods and/or new mathematical tools. 
Although the graph representation provides a convenient framework to search for MTLZ models, for large graphs, e.g. the descendants of $(0,2)$-graphs considered in Sec. VB, complete analysis seems difficult. 
We expect that new representations of MTLZ models beyond graphs, if they exist, could simplify the analysis and even render it analytically tractable. 
After all, the integrability conditions \eqref{int-cond} are constraints on matrices $H$ and $H'$, and for the MTLZ class these matrices are restricted to be linear in $t$ and $\tau$; these conditions form a set of commutation relations of matrices \cite{MTLZ-2020}, and they may be represented by some other mathematical objects which better reveal the underlying structures of these conditions. With those more advanced tools, hopefully the conjecture can finally be resolved. 

A possible extension of this work is to perform graph-theoretical searches for other classes of integrable time-dependent quantum models beyond the MTLZ class. In particular, there is another class of MLZ models named the $t/\tau$ family \cite{parallel-2020}, whose integrability can be utilized to achieve exact solvability. Models in this class have commuting partners of the form $H'=B_{12} t+B_{22} \tau+A_2+C/\tau$ where $C\ne 0$, namely, with a $1/\tau$ dependence. 
A common feature of this class of models is the existence of parallel diabatic energy levels; several classes of exactly solvable models, namely, the driven Tavis-Cummings model, the Demkov-Osherov model, and the generalized bow-tie model, all belong to this class. For the $t/\tau$ models, constraints from integrability can also be represented as data on graphs. Actually, in \cite{parallel-2020} all possible graphs up to $N = 6$ and some graphs with $N = 7,8$ are examined, and several new solvable models beyond the above mentioned classes are identified. As we did here for MTLZ models, for this $t/\tau$ class it would also be desirable to perform a general graph-theoretical analysis and to explore models on larger graphs, with the hope of finding new solvable models. In particular, one can follow a similar procedure as in the current work: first look for rules on graphs, and then identify minimal subgraphs based on these rules, and finally perform a graph search starting from the minimal subgraphs. A difference is that models in the $t/\tau$ family have parallel diabatic levels, so in a graph representation one needs to introduce an additional type of edges between states with parallel levels. We thus expect graph-theoretical analysis on the $t/\tau$ family to be more complicated than that on the MTLZ family.

Finally, we envision potential applications of discovering a new MTLZ model. In principle, such a solvable model can be useful when studying any practical physical systems where MLZ models can be realized. For example, it was theoretically proposed that MLZ models can describe transitions in quantum dot systems \cite{4-state-2002,Malla-2021}, and quantum control of qubits via Landau-Zener-St\"{u}ckelberg interference was experimentally achieved in these systems \cite{Cao-2013}. We thus expect quantum dot arrays to be promising platforms to realize MTLZ models: applying linear gate voltage sweeps on each dot, quantum states in a quantum dot array can be modeled theoretically by an MLZ model; couplings between dots can then be manipulated to realize parameters for an integrable MTLZ model. The exact solution of this model then provides an accurate description of transition probabilities in this quantum dot system, which can further guide the design of the array to implement desired quantum control or quantum state transfer protocols.

\section*{Acknowledgements}
This work was supported by the National Natural Science Foundation of China under Grant No. 12105094, and by the Fundamental Research Funds for the Central Universities from China.

\section*{Data availability}

The data that support the findings of this work are openly available \cite{github}.



\begin{thebibliography}{}
\bibitem{note-solvable} Exact Solvability of a multistate LZ model is usually defined as such that all its transition probabilities for an evolution from $t=-\infty$ to $t=\infty$ can be expressed in analytical forms.
\bibitem{footnote-complex-analogy} Note an interesting analogy: in complex analysis, the Cauchy-Riemann condition and the Cauchy integral theorem (freedom of deformation of integration contours in a complex plain without changing the complex integral) are differential and integral representations of the same property---analyticity, respectively. Here, the zero curvature condition \eqref{int-cond} and the freedom of path deformation in the $(t,\tau)$ plane \eqref{U-LZ-deform} are differential and integral representations of integrability, repsectively. The mathematical objects involved here are matrices instead of complex functions.
\bibitem{footnote-Cartesian} The Cartesian product $G \times H$ of graphs $G$ and $H$ is defined as a graph such that: its vertex set $V(G \times H)$ is the Cartesian product $V(G) \times V(H)$ (namely, $(u,v)\in V(G \times H)$  if and only if $u\in V(G) $ and  $v\in V(H) $); two vertices $(u,v)\sim (u' ,v' )$ in $G \times H$ if and only if either $u = u'$ and $v\sim v'$ in $H$, or $v = v'$ and $u\sim u'$ in $G$.
\bibitem{note-distance} The distance between two vertices in a graph is the length of a shortest path between the two vertices.
\bibitem{note-diameter} The diameter of a graph is the maximum distance between any two vertices in the graph. Note that by considering a finite diameter the graph is automatically a connected graph.
\bibitem{note-inversion} Note that for two sequences connected by inverting the numbers (e.g. $1341$ and $1431$), we only list one of them; the two cases are indeed equivalent when considering the $1221$-free case. However, when considering the case with $1221$, both sequences need to be considered, because the minimal subgraph for this case is not symmetric upon inversion of layers.
\end{thebibliography}
\end{document}